\documentclass{article}
\usepackage{arxiv}
\usepackage{graphicx}
\usepackage{float}
\usepackage{amsmath}
\usepackage{amssymb}
\usepackage{hyperref}
\usepackage{xcolor}
\usepackage{physics}
\usepackage{gensymb}
\usepackage{titlesec}
\usepackage{sectsty}
\usepackage{subcaption}
\usepackage{caption}
\usepackage{titletoc} 
\usepackage{etoolbox}
\usepackage{tocloft}
\usepackage{orcidlink}
\usepackage{bm}

\allsectionsfont{\centering}

\newcommand{\etal}{\emph{et al.}}

\title{Primordial black holes in Randall--Sundrum: Cosmological signatures}
\author{Itzi Aldecoa-Tamayo\,\orcidlink{0009-0002-1119-7198}\thanks{i.aldecoa-tamayo@sussex.ac.uk},
Christian T.~Byrnes\,\orcidlink{0000-0003-2583-6536}\thanks{c.byrnes@sussex.ac.uk},
and
David Seery\,\orcidlink{0000-0003-3421-6080}\thanks{D.Seery@sussex.ac.uk} 
\\[2ex]
\textit{Astronomy Centre, University of Sussex, Falmer, Brighton, BN1 9QH, UK} 
}
\date{}

\begin{document}
\vspace{-1.7em}
\maketitle
\begin{abstract}
    We reconsider primordial black hole physics in Randall--Sundrum Type-II
    universes,
    focusing on constraints from cosmological and astrophysical observables.
    We pay particular attention to
    scenarios
    that allow
    the entirety of dark matter to be in the form of
    higher-dimensional primordial black holes.
    This is possible for a range of AdS radii and black hole masses.
    Observable constraints are generally modified due to the
    changes in the higher-dimensional gravitational sector,
    and
    come from low-energy $e^{\pm}$ emission, microlensing, and
    possibly from contributions to unresolved radiation backgrounds.
    We discuss constraints from the cosmic microwave background
    due to injection of Hawking quanta into the
    intergalactic medium.
    Finally, we comment
    on recent discussions on the compatibility of higher-dimensional black holes and the KM3-230213A event.
\end{abstract}
\vspace{0.2cm}
\tableofcontents  
\newpage
\setcounter{footnote}{0}

\section{Introduction}
In the late 1990s, a number of higher-dimensional
cosmological models%
    \footnote{See Ref.~\cite{Rizzo} for a pedagogical introduction to
    higher-dimensional cosmological models.}
were introduced, motivated by string theory, unification of interactions,
and the search for solutions to the hierarchy problem.
In these frameworks, gravity is allowed to propagate in
additional spatial dimensions,
possibly
allowing the fundamental Planck scale to be reduced,
but also
changing gravitational dynamics at small scales.
Many such models have been proposed,
distinguished by features such as the number of extra dimensions,
their curvature, and whether the higher-dimensional space is finite or infinite.
A notable example is provided by braneworld scenarios,
in which our observable Universe is described as a lower-dimensional
brane embedded in a higher-dimensional bulk. In this setting, Standard Model
fields are confined to the brane,
while gravity propagates into the higher-dimensional bulk.
This modifies both cosmological dynamics and black hole physics at small scales.
Scenarios of this kind remain phenomenologically relevant and often exhibit
distinctive features in early epochs.

Primordial black holes%
    \footnote{For extensive reviews on primordial black holes,
    see for example Refs.~\cite{review-pbh-1,review-pbh-2,review-pbh-3}.}
(``PBHs''),
hypothesized to form in the early Universe \cite{PBH-Hawking},
provide a natural probe of these scenarios.
Multiple mechanisms have been proposed for their formation,
including the collapse of overdense regions seeded by primordial fluctuations,
bubble collisions during phase transitions,
and the dynamics of topological defects.
Interest in PBHs has been increased by the possibility that they could
constitute a fraction (or even all) of the dark matter (DM).
Unlike astrophysical black holes, PBHs could span a wide range of masses,
and their evaporation or survival until the present epoch would reflect
the gravitational laws governing their dynamics. Therefore,
their formation and
evolution leave observable signatures,
offering a way to test gravity in regimes inaccessible
to laboratory experiments.

Cosmological signatures of PBHs have been studied
in the
``large extra dimensions" (``LED'') scenario,
where the higher-dimensional space consists of $n>2$ flat and compact
dimensions \cite{LED}.
These studies
(see Refs.~\cite{Friedlander-I,Friedlander-II,Johnson}) suggested
significant modifications to black hole evaporation and
cosmological constraints. In the Randall--Sundrum (RS)
braneworld \cite{RS-I,RS-II}, where there is a single extra dimension
with anti-de Sitter (AdS) curvature, similar investigations were
carried out in the early 2000s, primarily through analytic estimates.
These analyses established many of the qualitative features of
PBH behaviour in warped geometries---%
covering their formation and evaporative evolution \cite{Clancy-I},
accretion \cite{Clancy-II,Majumdar}, and astrophysical
constraints \cite{Clancy-III}---%
and provided the foundation for subsequent developments.
Building on this groundwork, we revisit the problem by performing
a more detailed analysis of the evaporation process and
interaction with the cosmological background, making use of public
computational tools developed in the intervening years,
and by incorporating updated cosmological and astrophysical constraints
from more refined observations.
In particular, we focus on Randall--Sundrum single-brane models,
the so-called Type-II scenarios~\cite{RS-II},
with the primary aim of defining the observationally allowed
window for PBH dark matter.

The structure of this article is as follows.
In Section \ref{section-PBH-BW} we briefly review
higher-dimensional brane cosmology scenarios,
and also the physical properties
of higher-dimensional black holes.%
    \footnote{For a complete picture we recommend the
    reviews by Maartens~\cite{Marteens} and
    Emparan~\cite{Emparan}, respectively.}   
In Section~\ref{section-evap}
we give a brief discussion of Hawking evaporation in braneworlds.
In Section \ref{section-EM-radiation} we cover mechanisms
related to black hole evaporation (directly or indirectly)
that produce electromagnetic radiation, and discuss the expected flux.
We then focus on the detection of low-energy electrons and positrons
in the Galaxy (Section \ref{section-voyager}).
In Section \ref{section-CMB} we discuss the impact of energy injection
on the cosmic microwave background (CMB), sourced by evaporating black holes.
In Section \ref{section-microlensing} we give a short discussion
of microlensing in braneworld scenarios.
For each observable, we consider cosmological
observations and obtain corresponding bounds
on the primordial black hole abundance.
Finally, we
comment on recent work relating higher-dimensional black holes
to the KM3-230213A event.
We conclude with an overview of our study
by showing aggregate constraints on the PBH abundance
in Section \ref{section-conclusions}. 

\textbf{Notation}. We use natural units $c=\hbar=k_B=1$.
However, when discussing observables and detections,
we convert the results back to physical units \cite{eemeli}.
%
%
%
%
\section{Primordial black holes in braneworlds}\label{section-PBH-BW}
\subsection{Early Universe in the brane}
In the
RS-II scenario,
our universe is a $(3+1)$-dimensional brane embedded in an
otherwise empty $(4+1)$-dimensional AdS bulk,
taken to be $\mathbb{Z}_2$
symmetric about the brane. The bulk cosmological constant is
\begin{equation}
    \Lambda_5=-\frac{6}{l^2}\,,
\end{equation}
where $l$ is the curvature radius, and the AdS$_5$ metric takes the form
\begin{equation}\label{RS-metric}
    ^{(5)}\dd s^2=e^{-2|y|/l}g_{\mu\nu}\dd x^\mu\dd x^\nu+\dd y^2
\end{equation}
in Gaussian normal coordinates,%
    \footnote{In this choice of coordinates $x^A=(x^\mu,y)$, $y$
    is a coordinate orthogonal to the brane ($n_A\dd x^A=\dd y$,
    with $n^A$ the unit normal) and the brane is located at $y=0$.}
with $g_{\mu\nu}$ being the induced metric on the brane.
The brane is located at $y=0$.
The AdS radius $l$ characterizes the $y$-distance at which the
approximation of a flat fifth dimension breaks down.
It also sets the effective Planck scale on the brane, $M_4$, as
a function of the fundamental Planck scale $M_5$,
\begin{equation}
    \frac{1}{l^2}=\frac{M_5^6}{M_4^4}-\frac{\Lambda_4}{3}
\end{equation}
with $\Lambda_4$ the 4D cosmological constant and $M_4$
the effective Planck mass on the brane. The value of $l$
is constrained by table-top experiments, 
which require $l\lesssim10^{-6}$m \cite{grav-tests}.

Dynamics on the brane are affected by the presence of the extra
dimension. In particular, the effective $(3+1)$-dimensional Einstein equations
are modified. This results in the following Friedmann equation on the brane,
\begin{equation}\label{Friedmann}
    H^2=\frac{8\pi}{3M_4^2}\left(\rho+\frac{\rho^2}{2\lambda}+\rho_\mathcal{E}\right)+\frac{\Lambda_4}{3}-\frac{k}{a^2}
\end{equation}
where $H$ is the Hubble parameter on the brane ($H=\dot{a}/a$), $a$ is the scale factor in brane coordinates,
$\rho$ is the energy density of ordinary matter
fields on the brane, and $k=+1, -1, 0$
for a closed, open or flat Friedmann geometry on the brane, respectively. 
The effective Friedmann equation~\eqref{Friedmann} contains
two additional terms sourced by
gravitational backreaction and dimensional reduction of the complete geometry:
\begin{itemize}
    \item The quadratic term $\rho^2/2\lambda$, 
    where $\lambda$ is the brane tension, a parameter that depends purely on the fundamental scales:
    \begin{equation}
        \lambda=\frac{3M_5^6}{4\pi M_4^2}\,.
    \end{equation} 
    \item The dark radiation $\rho_\mathcal{E}\sim a^{-4}$, resulting from a non-local response to disturbances in the bulk.
    It is expected to be non-vanishing in the vicinity of inhomogeneities,
    but is constrained to be very small at the level of the background; e.g.,~$\rho_\mathcal{E}/\rho_{\rm rad}\lesssim 0.062$ at the epoch of Big Bang Nucleosynthesis \cite{KK-BBN-1,KK-BBN-2}. 
\end{itemize}
For early epochs (i.e.~density parameter $\Omega_{\Lambda_4}\ll 1$)
on a flat Friedmann brane ($k=0$), Equation~\eqref{Friedmann} can be approximated as
\begin{equation}
    H^2\approx\frac{8\pi}{3M_4^2}\left(\rho+\frac{\rho^2}{2\lambda}\right)\,.
\end{equation}
One sees immediately that, depending on the value of $\rho$,
the Universe will follow an expansion history
dictated by the standard relation $H^2\sim\rho$
(which we describe as the ``linear'' regime),
or by the non-conventional scaling relation $H\sim\rho$
(which we describe as the ``quadratic'' regime).
The latter occurs in the very early Universe when $\rho\gg\lambda$.
The crossover between regimes occurs approximately when $\rho \sim \lambda$,
at a cosmic time $t_c$,
\begin{equation}
    t_c\equiv \frac{l}{2} .
\end{equation}
At early times $t \lesssim t_c$,
the energy density is large
and we are in the quadratic regime.
Assuming a radiation background with $\rho \propto a^{-4}$,
we have
\begin{equation}\label{quad-reg}
    a=a_0\left(\frac{t}{t_0}\right)^{1/4}\,,\qquad    \rho=\frac{3M_4^2}{32\pi t_c t}\,,\qquad  H^{-1}=4t .
\end{equation}
Meanwhile,
for $t \gtrsim t_c$
the standard cosmology is recovered,
\begin{equation}
      a=a_0\left(\frac{t}{t_0^{1/2}t_c^{1/2}}\right)^{1/2}\,,\qquad  \rho=\frac{3M_4^2}{32\pi t^2}\,,\qquad     H^{-1}=2t\,.
\end{equation}
\subsection{Black hole metrics}
\label{sec:bh-metrics}
The effects of higher-dimensional geometry
can modify both the physics of gravitational collapse,
and the compact objects that form at its endpoint.
In particular,
$(3+1)$-dimensional black hole notions such as uniqueness may not extend to higher-dimensional scenarios.%
    \footnote{In $(4+1)$ dimensions,
    Myers--Perry black holes and black rings are both solutions
    to the Einstein equations and prove that uniqueness is violated.
    We consider only black holes in this work, since that is what one expects
    from spherical collapse of brane fields in an otherwise empty bulk.}
Also, compact or warped extra dimensions may prevent a spherical topology
for the event horizon~\cite{Emparan}.

Consider black holes in RS-II.
If the horizon radius
is significantly smaller than the AdS radius ($r_0\ll l$),
compact objects are not sensitive to the
warping of the extra dimension.
Therefore, \emph{in the near-horizon limit},
one can rely on higher-dimensional generalizations of the
Schwarzschild (or Kerr) metrics. Black holes in this category have relatively low masses. We describe them as \emph{small} black holes.

The higher-dimensional generalization of the
Schwarzschild metric is known as the Schwarzschild--Tangherlini
solution \cite{ST-metric}. In $(4+1)$-dimensions it is
\begin{equation}\label{ST-metric}
     ^{(\text{ST})}\dd s^2=-\left(1-\frac{r_0^2}{r^2}\right)\dd t^2+\left(1-\frac{r_0^2}{r^2}\right)^{-1}\dd r^2+r^2\dd\Omega_3^2
\end{equation}
where $\dd \Omega_3$ is the
element of area of a unit $3$-sphere, 
and $r_0$ is the radius of the black hole event horizon,
\begin{equation}\label{horizonradius}
    r_0=\sqrt{\frac{8}{3\pi}}\frac{\sqrt{lM}}{M_4}\,,
\end{equation}
where $M$ is the black hole mass.

In this paper we carry out the analysis from the perspective of a 
brane observer.
In Gaussian normal coordinates (see Equation \eqref{RS-metric}),
the induced metric on the brane $g_{\mu\nu}$
is simply \eqref{ST-metric},
with the replacement
$\dd \Omega_3^2\rightarrow\dd\Omega_2^2$.

Note that the brane tension
means that the status of brane-based black holes is somewhat different
in the RS-II framework, compared to alternative scenarios
such as the Arkani-Hamed {\etal} LED models.
As shown by Fraser and Eardley \cite{fraser},
small black holes on a positive‑tension RS‑II brane
exhibit substantial gravitational binding energy,
making them stable against escape into the bulk.
Meanwhile,
in LED models (i.e.~no brane tension),
small black holes localized on the brane are not generically bound and can escape into the bulk via recoil from asymmetric
Hawking emission into bulk modes \cite{frolov-1,frolov-2,flachi}.
Moreover, for higher-codimension models,
the warping induced in the bulk is no longer AdS$_5$,
and its geometry exhibits a deficit in the measure of the
transverse angular space \cite{vilenkin}.
This
affects the small black hole metric.
For example, for codimension-2 branes,
embedded small black holes retain the local radial structure of the
Schwarzschild–-Tangherlini solution.
However, their total horizon area is reduced
in proportion to the missing angular measure \cite{kaloper},
with corresponding changes in their thermodynamic properties.
The RS-II model is thus a special case in the space of
higher-dimensional frameworks,
where small black holes preserve the metric
\eqref{ST-metric} and are naturally bound to the brane.

Conversely,
in the case $r_0\gg l$, the transverse extension of the black 
horizon into the
extra dimension is negligible in comparison to its radial extension
along the brane directions.
Therefore,
at least to obtain the geometry on the brane,
one can use the $(3+1)$-dimensional Schwarzschild
or Kerr metrics as an
approximation. Black holes in this category have relatively high masses,
and we describe them as \emph{large} black holes. 
The transition between these two regimes is still
poorly understood. The full
description of a brane-based black hole is expected to have a
\textit{flattened pancake} shape \cite{emparan-2}
with a non-negligible thickness.
To date, no exact analytical solution for the spacetime is known.

However, neither of these near–horizon descriptions controls the far–field regime, where the gravitational field is weak and linearized gravity applies. In this asymptotic limit, the Garriga--Tanaka analysis~\cite{Garriga--Tanaka} shows that the potential acquires a universal correction of order $1/r^3$, reflecting the influence of the bulk extra dimension and gravity \textit{leaking} into the extra dimension. The solution that captures the correct asymptotics for both small and large brane black holes reads
\begin{equation}\label{GT-metric}
   ^{(\text{GT})} \dd s^2=-\left(1-\frac{2M}{r} -\frac{4Ml^2}{3r^3}\right)\dd t^2
    +\left(1+\frac{2M}{r} +\frac{2Ml^2}{3r^3}\right)\dd r^2+r^2\dd \Omega_2^2\,.
\end{equation}
It is clear that \eqref{ST-metric} does not approach
\eqref{GT-metric} in the limit $r/r_0 \gg 1$.
This poses a problem.
There are separate solutions for the near-horizon and asymptotic regions,
but there is currently no analytic solution that smoothly interpolates
between
the two.
In this paper, we employ the
Schwarzschild--Tangherlini solution for near-horizon processes, such as Hawking evaporation (discussed in Sections \ref{section-EM-radiation}-\ref{section-CMB}), and the Garriga--Tanaka solution for far-field processes, such as microlensing (discussed in Section \ref{section-microlensing}). 
\subsection{Primordial black hole population}
Constraints on the PBH population depend on the
mass function
we impose.
To make the discussion as clear as possible, we work under a set of simplifying assumptions, which we detail below. 

We consider the PBH population to lie either in the ``small" or ``large" black hole regime. In both cases, we assume a monochromatic mass spectrum with approximately constant mass until today. For the ``small" black holes, we additionally assume that their five-dimensional geometry is preserved throughout their lifetime.

\par\noindent\textbf{Large black holes.}
First consider ``large'' black holes, in the sense
of Section~\ref{sec:bh-metrics}.
These are black holes whose mass is sufficiently
large
that the on-brane metric can be
approximated as $(3+1)$-dimensional
Schwarzschild or Kerr.

In the conventional formation scenario, 
PBHs are produced by direct collapse of large-amplitude
density perturbations as they re-enter the horizon.
Since the collapse process
is relatively rapid,
almost the entire horizon volume is expected
to undergo monolithic collapse to a single black hole.
The horizon mass grows with time,
so perturbations entering the horizon later
form more massive black holes.
For black holes forming at times $t \gtrsim t_c$,
we find that the horizon mass is always
sufficient to make the black hole ``large''.

Black holes in this regime behave
in the same way as the standard cosmology,
and experience the same expansion history.

\begin{figure}[htbp]
    \centering
\includegraphics[width=0.5\linewidth]{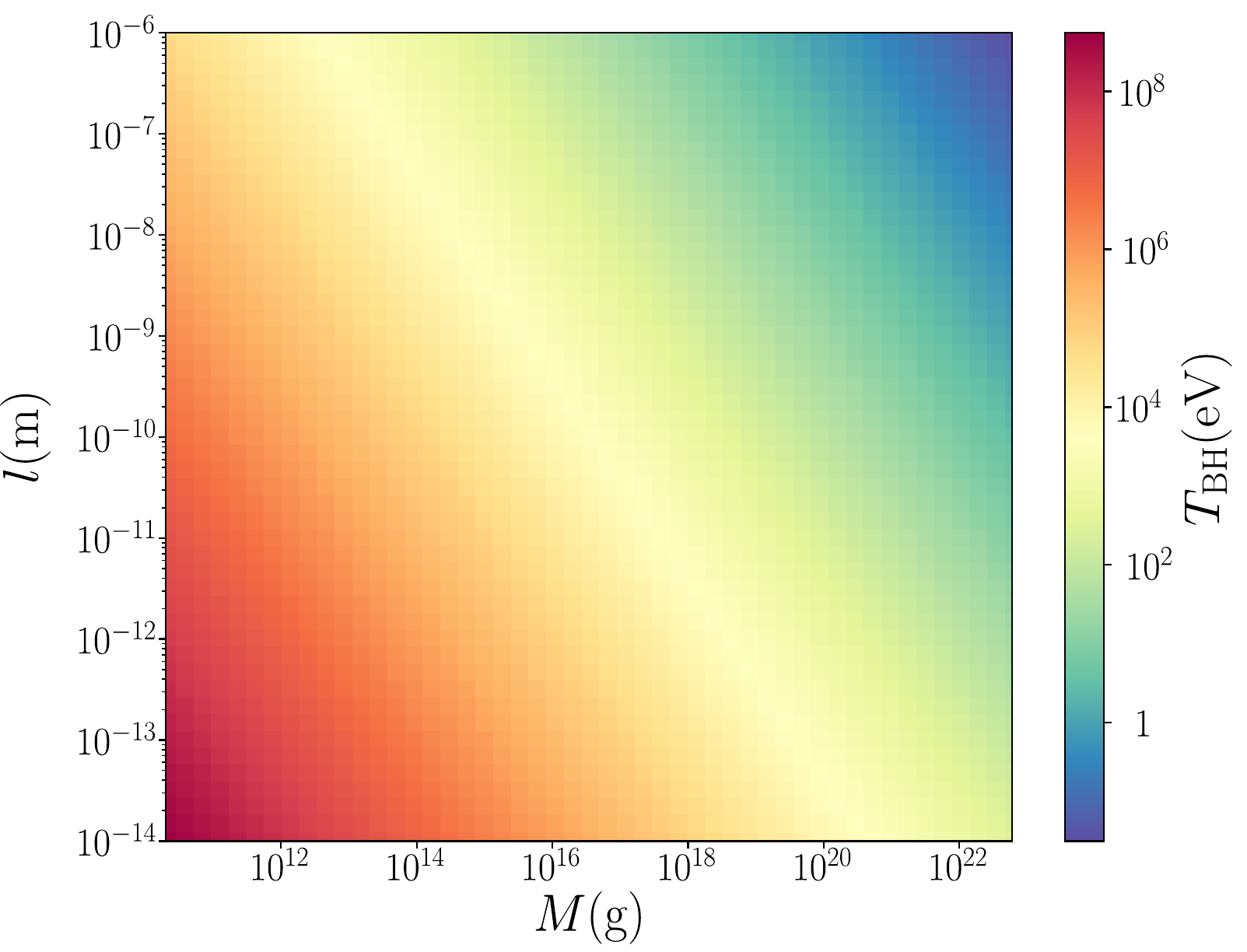}
    \caption{Black hole temperature dependence on the black hole mass $M$ (in grams) and the bulk AdS radius $l$ (in metres)
    for Schwarzschild--Tangherlini black holes in Randall--Sundrum Type-II universes, following \eqref{horizonradius} and \eqref{st-temp}.}
    \label{fig-temperature}
\end{figure}

\par\noindent\textbf{Small black holes.}
Now consider black holes forming at times
earlier than the crossover $t_c$.
At this time, the horizon scale is of order
$H^{-1} \sim l$.
It can be checked that this implies
such black holes form in the ``small'' regime.
In principle, this leaves open the question of
whether they accrete sufficiently to move into 
the ``large'' regime before $t_c$.
However, the event horizon of any black hole
must fit within the cosmological horizon,
and, therefore, its radius is also bounded by $l$.
The conclusion is that,
irrespective of the unknown details of collapse and accretion
for $t \lesssim t_c$,
all black holes
formed prior to $t_c$ remain in the ``small''
regime.%
    \footnote{Black hole evolution in the
    quadratic regime $t < t_c$
    was discussed in detail by
    Clancy {\etal}~\cite{Clancy-I,Clancy-II,Clancy-III}
    and Majumdar~\cite{Majumdar}.
    We intend to revisit this analysis in future work.}

We set initial conditions corresponding to a specified mass
at $t = t_c$, and track their subsequent evolution.
This is sufficient for the purposes of the analysis in this paper,
but would be inadequate if we wished to relate PBH
masses to specific $k$-modes in the primordial power spectrum.

Black holes in the ``small'' regime have a metric
given by the restriction
of Schwarzschild--Tangherlini
(or Myers--Perry)
to the brane surface $y=0$.
In particular, their mass-to-radius
and mass-to-temperature relations are
modified
compared to black holes in $(3+1)$ dimensions.
The temperature relation is
\begin{equation}\label{st-temp}
    ^{(\text{ST})}T_{\text{BH}}
    =
    \frac{1}{2\pi r_0} 
    ,
\end{equation}
where $r_0$ is the horizon radius. They also emit a modified spectrum
of Hawking radiation,
to be discussed in Section~\ref{section-evap} below. The dependence of the black hole temperature on the mass and AdS curvature can be seen in Figure \ref{fig-temperature}.

\par\noindent\textbf{Mass evolution.}
For both ``large''
and ``small''
black holes, we take their
mass to be approximately constant throughout
their lifetime.
In both cases,
accretion is expected to enhance
their masses only by an $\mathcal{O}(1)$ factor for $t>t_c$. See details of this argument for the case of a RS-II
cosmology in Ref.~\cite{Clancy-II}. Given other uncertainties in the calculation, this barely
affects estimates of their lifetime.

We also neglect mass loss
due to emission of Hawking quanta, except in the final
stages of evaporation. This is a standard
approximation for large black holes.
Let us consider why it is also a reasonable approximation
for most PBH masses in the small black hole limit.
For a fixed mass, Schwarzschild--Tangherlini black holes
have Hawking temperature given by Equation~\eqref{st-temp}.
The total emission rate
into brane degrees of freedom
(which is expected to dominate over emission
into the bulk \cite{Emparan-brane}) is proportional to
\begin{equation}\label{ev-scaling}
    \frac{dM}{dt}\propto A_{\rm eff} T^4 \propto T^2 \propto M^{-1}\,,
\end{equation} 
with the
effective area $A_{\rm eff}=4\pi r_{\rm eff}^2$ and $r_{\rm eff}=2r_0$ \cite{Emparan}. Integrating \eqref{ev-scaling}
tells us that the lifetime $\tau$ of a black hole with
initial mass $M_i$ can be written
\begin{equation}
    \tau =k M_i^2,
\end{equation}
where $k$ an approximate constant,
assuming that the number of light degrees of freedom
available for Hawking emission
does not vary significantly over the lifetime
of the black hole.
We define $M_*$ as the black hole mass whose lifetime
is equal to the current age of the Universe, i.e.~$t_0=kM_{*}^2$.
Now, consider a black hole with initial mass larger than $M_*$,
i.e., $M_i=\psi M_*$, with $\psi>1$.
The mass of this black hole today $M_0$ satisfies
\begin{equation}
    t_0=k(M_i^2-M_0^2)\,,
\end{equation}
and therefore
\begin{equation}
    M_0=M_*\sqrt{\psi^2-1}\,.
\end{equation}
We conclude that
the change in mass, due to evaporation up to the present day
for a black hole with initial mass $M_i\geq 3M_*$,
is less than about $6\%$.
In what follows,
we assume a constant mass for all black
holes with $M_i\geq 3M_*=M_\text{min}$.

\par\noindent\textbf{Monochromaticity.} Finally, we comment on the 
mass spectrum. While most conventional PBH formation scenarios motivate an approximately monochromatic spectrum, we take it to be exactly monochromatic to minimize parameters and facilitate a clearer conceptual analysis. Note, however, that the PBH dark matter fraction,
\begin{equation}\label{fpbh-1}
f_{\text{PBH}}\equiv\frac{\Omega_{\text{PBH},0}}{\Omega_{\text{DM},0}}\,,
\end{equation}
where $\Omega_{i,0}$ is the density parameter today of species $i$, is constrained more strongly for extended spectra in the case of standard (3+1)-dimensional Hawking radiation \cite{monochromatic-or-not}.
This is a result of $f_{\rm PBH}$ scaling steeply with the black hole mass on the low mass end of the asteroid mass gap, approximately following $f_{\rm PBH}\sim M^3$. Even for PBHs originating from a narrow peak in the primordial power spectrum, the asteroid-mass window tightens by roughly half an order of magnitude in terms of the minimum mass \cite{Gorton:2024cdm}. For five-dimensional black holes, the low-mass tail scaling is less steep but still prominent, close to $\sim M^2$. Therefore, we expect the asteroid mass window to shrink when going beyond the assumption of a monochromatic mass function, albeit by a smaller amount in braneworld models than in the conventional scenario. 
%
%
%
%
\section{Hawking evaporation in braneworlds}\label{section-evap}
Hawking evaporation in both the ``large'' and ``small''
regimes of RS-II has been a focus of debate.
Heuristic arguments have progressively led to a
consensus on their evaporation dynamics, which has been substantiated by numerical analyses. In what follows, we present a brief overview of the key developments that have shaped this field.

Evaporation depends crucially on the relative size of the
horizon and the AdS radius.
In principle, two distinct radiative channels are available.
First, Hawking quanta may be emitted into
brane matter fields.
These are intrinsically four-dimensional, and are restricted to the brane.
Second,
quanta may be radiated into bulk
degrees of freedom.
These are
higher-dimensional gravitons in the case of an otherwise empty bulk.
Such bulk modes propagate in all five dimensions.
It is not immediately clear which channel is the dominant
emission
mechanism,
and
addressing the balance requires a careful analysis of the underlying dynamics. 

\par\noindent\textbf{Large black holes.}
In this case, the bulk channel admits a
dual holographic interpretation, corresponding
to coupling the black hole to a large number
of strongly interacting conformal fields on the brane.
This would suggest
the black hole has access to many additional
degrees of freedom~\cite{emparan-3}.
For this reason
it was originally proposed that large black holes would evaporate much more rapidly than in four dimensions.%
    \footnote{Note that in LED frameworks such holographic
    considerations do not apply,
    and large black holes are understood to evaporate
    according to the standard four-dimensional
    law once their horizon exceeds the compactification
    scale~\cite{Emparan-brane}.}
If so, the observed persistence of long-lived astrophysical
black holes would impose an upper bound on the AdS radius,
since otherwise their enhanced evaporation would have already depleted them. 

This is no longer believed to be the case.
Wiseman {\etal}~\cite{wiseman} argued that the naïve $N^2$-enhanced Hawking
flux was not generic: static, localized brane black holes 
need not radiate into the full tower of holographic modes.
This was later confirmed by the numerical construction of large,
static RS-II black holes \cite{figueras},
which demonstrated that long-lived solutions indeed exist.
Most recently, Emparan {\etal}~\cite{emparan-4} clarified
that the enhanced evaporation is restricted to the
``connected'' (funnel) phase,
where the brane horizon merges with a bulk horizon
and energy can flow into the bulk.%
    \footnote{This argument also applies in the ``small'' black hole limit.}
Conversely,
in the ``disconnected'' (droplet) phase,
which provides a viable branch of solutions,
the black hole is effectively insulated and evaporates
essentially as in four dimensions,
up to small corrections. In this case,
the phenomenology associated to Hawking evaporation of
large black holes in RS-II becomes observationally
indistinguishable from that in LED scenarios. 

\par\noindent\textbf{Small black holes.}
Holographic arguments do not apply to small black holes.
Provided they are also ``disconnected", these radiate primarily
into brane degrees of freedom,
with only a small fraction of the flux into the bulk
graviton modes~\cite{Emparan-brane}.
Their evaporation proceeds via the generalized Hawking process
in five dimensions, and is thus essentially the same as in LED frameworks.
However, unlike RS-II,
five-dimensional LED scenarios are effectively ruled
out by LHC-based evidence combined with the bounds imposed by
table-top experiments.

\par\noindent\textbf{Greybody factors.}
For a black hole with zero
angular momentum ($J=0$), the rate of emission
of Hawking quanta
a
massless
on-brane state of spin $s$,
with energy between $E$ and $E + \dd E$,
can be written~\cite{Page-I}
\begin{equation}
    \label{ddotN}
    \frac{\dd \dot{N}^{\mathrm{P}}_s}{\dd t}
    =
    \frac{\dd E}{2\pi}
    \frac{\Gamma_s(\tilde{\omega})}
    {e^{E/T_{\text{BH}}} - (-1)^s} .
\end{equation}
Here,
$\text{P}$
stands for primary (i.e.~direct) emission,
and $\Gamma_s$
is the greybody factor.
It is responsible
for departures from an exact thermal spectrum.
It depends
on the number of extra dimensions $n$,
and any quantum numbers carried by the mode.
It also depends on $E$ via the dimensionless
combination $\tilde{\omega} \equiv r_0 E \sim E / T_{\text{BH}}$.
It is computed by solving an
effective four-dimensional wave equation
for a set of 
states labelled by angular momentum
quantum numbers $\ell$, $m$, and summing over these labels.
The corresponding mass loss rate is
given by summing over species $j$,
\begin{equation}
    -
    \frac{\dd M}{\dd t}
    =
    \frac{\dd E}{2\pi}
    \sum_{j}
    g_j
    \frac{E \, \Gamma_j(\tilde{\omega})}
    {e^{E/T_{\text{BH}}} - (-1)^s} ,
\end{equation}
where $g_j$ counts the number
of polarizations (or helicities)
of species $j$.
Equation~\eqref{ddotN}
shows that,
in a RS-II scenario,
Hawking emission
for ``small'' black holes
is modified in two ways.
First,
for a fixed mass $M$,
such black holes are intrinsically
colder because they satisfy five-dimensional
thermodynamic relations.
This makes their horizon radius larger, and their temperature
lower.
Second,
the details of the greybody factors $\Gamma_s$
are modified.
``Large'' black holes are described by the
$(3+1)$-dimensional Schwarzschild geometry,
so the mass--temperature relation and greybody factor
revert to their standard values.

The modified greybody factors for ``small'' black holes were
calculated numerically
for non-rotating black holes by Harris \& Kanti~\cite{kanti-1}.
The same authors extended
the formalism to
emission of scalar modes with
nonzero
angular black hole
momentum $J \geq 0$~\cite{kanti-2}.
The extension to
gauge fields was given by Casals {\etal}~\cite{kanti-3}.
Later, Ida {\etal}~extended the calculation
to fermions \cite{Ida}.
Both groups confirmed the heuristic argument
that higher-dimensional black holes
emit mainly on-brane modes,
given in Emparan {\etal}~\cite{Emparan-brane}.

To compute emission rates, we use
fitting functions for
the greybody factors
extracted from the
computational tool \textsc{BlackMax}~\cite{BlackMax}.
We use these rates to 
determine the lifetime and temperature of
a black hole of given initial mass.
This enables us to determine the rate at which Hawking
quanta are injected into the Universe.
In practice, we find that lifetimes and emission
rates are hardly altered by inclusion
of the correct greybody factors,
compared to exact blackbody emission,
or if the massless approximation is dropped.

Equation~\eqref{ddotN}
has an important implication for
observational signatures. In this framework,
the Hawking spectrum of an individual black
hole—and therefore its cosmological imprint
through energy injection—depends solely on its temperature,
independent of the black hole mass $M$
or AdS radius $l$.
As a result, evaporative bounds on a primordial black hole population
constrain
only the number density
$n_{\text{PBH}} = f_{\text{PBH}}/M$
at fixed temperature.
The result is that
constraints on $f_{\text{PBH}}$
become
\emph{stronger}
at larger AdS radii.
Equations~\eqref{horizonradius}
and~\eqref{st-temp}
show that, for fixed $T_{\text{BH}}$
and $M_4$,
the black hole mass scales like $l^{-1/2}$.
Therefore the mass needed to achieve a specified
temperature $T_{\text{BH}}$ becomes smaller
as $l$ increases.
It follows that, for two AdS radii $l_1$ and $l_2$,
the constraints on $f_{\text{PBH}}$
are related via
\begin{equation}
    f_{\text{PBH}}(l_1)
    =
    \left(
        \frac{M(T,l_1)}{M(T,l_2)}
    \right)
    \,
    f_{\text{PBH}}(l_2)
    =
    \left(
        \frac{l_2}{l_1}
    \right)^{1/2}
    \,
    f_{\text{PBH}}(l_2)
    .
\end{equation}

This interplay, while not readily apparent in the usual
$f_{\text{PBH}}$--$M$
plots, allows for a useful translation between different
parameter choices.

\par\noindent\textbf{Transition to ``small'' regime.}
Note that as ``large'' black holes evaporate,
their surface area shrinks.
Eventually, the horizon radius $r_0$ becomes comparable to the AdS
radius $l$. In this regime, the geometry is expected to
transition between the ``large'' and ``small''
regimes.

In the absence of an exact brane-black-hole solution describing
this intermediate configuration and its time dependence,
the detailed evaporation law and the approach to a
spherically symmetric five-dimensional configuration
remain uncertain. It is reasonable to expect,
however, that an approximate interpolation between the
``large'' and ``small''
black hole phases should yield suitable results.
Non-exact numerical solutions (e.g.~\cite{figueras}, \cite{tanaka}
and \cite{santos}) should also be good enough for a PBH lifetime modeler.

In a cosmological context, in the absence of unusual accretion, these black holes must have formed shortly after $t_c$.
The AdS radius must be very small ($l\lesssim 10^{-14}$m)
in order for ``large'' black holes to have evaporated sufficiently
to find themselves at this crossover today.
This can be visualized by shifting the shaded regions
in Figure \ref{emission-allowed} to the right as the Universe ages. 
%
%
%
%
\section{Electromagnetic radiation}\label{section-EM-radiation}
\subsection{Production mechanisms}
\label{sec:em-production}
The total photon flux stemming from Hawking evaporation has
contributions from several production channels.
Direct emission is always present.
Secondary emission can occur
as a result of in-vacuo QED processes, involving
emitted quanta,
that produce final state photons.
Alternatively, it may occur due to
interactions between emitted quanta
and a medium.
Examples include
(non-)relativistic positron annihilation,
positronium formation and decay,
in-medium bremsstrahlung
and inverse Compton scattering.

For each process we require the instantaneous rate
for
Hawking emission
of the particles involved, obtained
from Equation \eqref{ddotN}.
In the following sections,
we discuss the channels that contribute most significantly
to the total photon flux.
We do not include processes with negligible impact.
For instance, secondary emission via hadron and gauge boson decay
is taken to be negligible. This is because primordial black holes
in the parameter space under consideration have temperatures
significantly lower than the QCD scale.
Therefore, hadron and gauge boson emission are expected to be
highly suppressed.%
    \footnote{In $(3+1)$-dimensional spacetime,
    the black hole temperature at which secondary photons
    start to dominate is 50 MeV \cite{Carr-2020},
    which is around $\Lambda_{\text{QCD}}/6$.
    We cannot know exactly when this transition occurs
    in RS-II without comparing the primary and secondary spectra
    to find the temperature at which the flux is comparable.
    However, we expect
    it should not differ by orders of magnitude.}
    
In Ref.~\cite{medium-brem} it was shown that,
for positrons with $E_{e^+}\lesssim 100\,\text{MeV}$,
in-medium bremsstrahlung and inverse Compton energy emission
can be neglected. Except for the final bursts in their evaporation,
braneworld black holes are generally too cold for these
production channels to be significant.

For $M\geq M_{\text{min}}$, the total flux can be approximated
\begin{equation}\label{sum-prods}
    \frac{\dd\dot{N}_\gamma^{\text{tot}}}{\dd E_{\gamma}}\approx  \frac{\dd \dot{N}^\text{P}_\gamma}{\dd E_\gamma}+\frac{\dd \dot{N}^\text{FSR}_\gamma}{\dd E_\gamma}+\frac{\dd\dot{N}^\text{IA}_\gamma}{\dd E_\gamma }+\frac{\dd\dot{N}^\text{NR}_\gamma}{\dd E_\gamma}\,,
\end{equation}
where ``FSR''
includes final state radiation from all relevant charged
particles emitted by the black hole;
``IA'' stands for inflight annihilation of relativistic positrons;
and ``NR'' includes all non-relativistic processes that take place
after positrons thermalise with the interstellar medium (ISM).

\begin{figure}[htbp]
    \centering
    \includegraphics[width=0.5\linewidth]{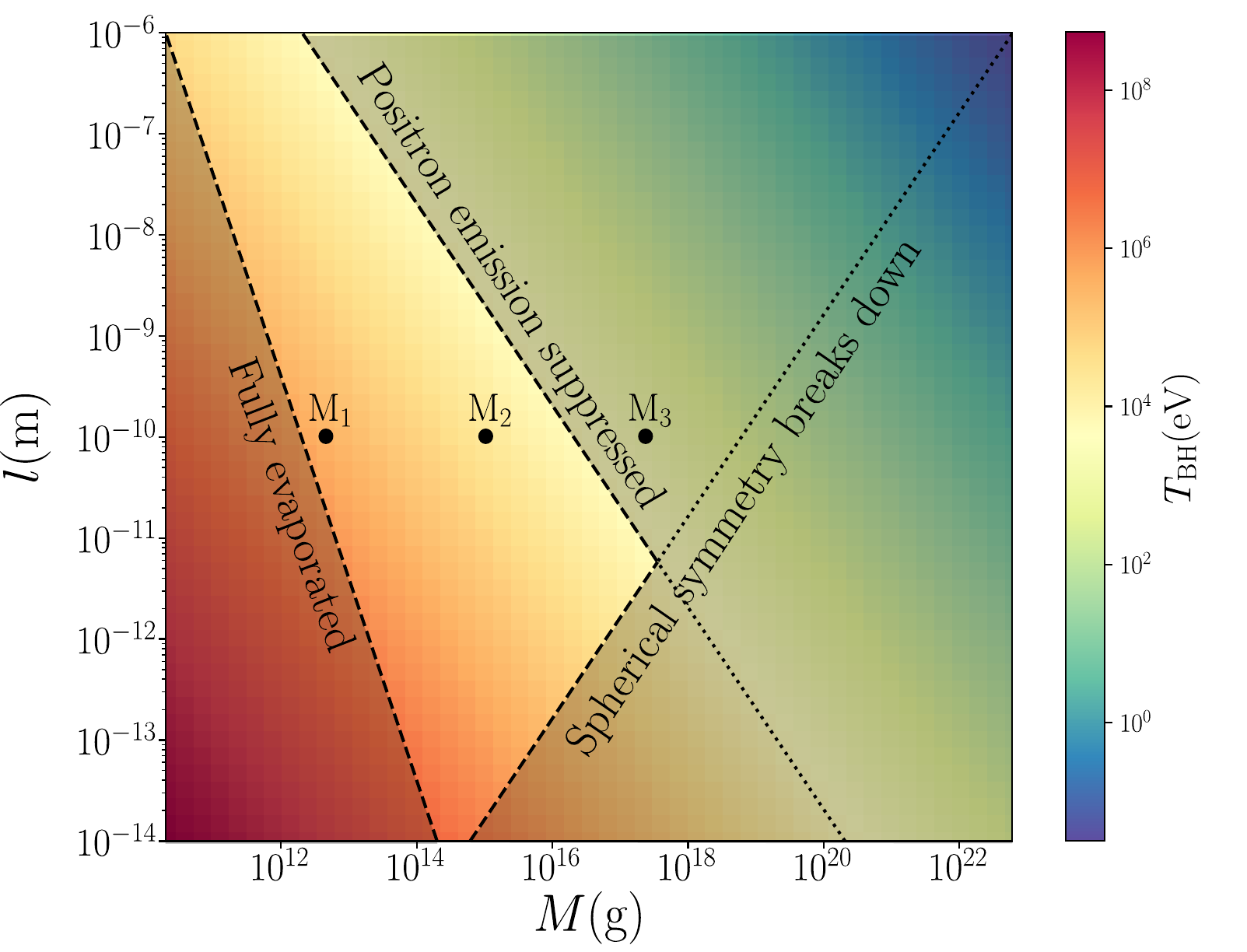}
    \caption{Dependence of positron emission on the black hole temperature. In the unshaded region of the parameter space $(M,l)$ positron/electron (or 511~keV photon) emission is favoured. The masses $M_1$, $M_2$ and $M_3$ for AdS radius $10^{-10}$m are used in Section \ref{section-EM-radiation}.\ref{photon-distribution} for the study of the extragalactic photon flux, see Figure~\ref{EGB-plots}.}
    \label{emission-allowed}
\end{figure}

Since some of the channels listed above require the emission of
massive particles, their contribution to the total flux depends
strongly on the temperature of the black hole. The range of masses
where $e^\pm$ emission is not suppressed is shown in
Figure \ref{emission-allowed}.
\subsubsection*{Direct Hawking emission}
The reduced temperature of higher-dimensional black holes, in contrast to the four-dimensional case, leads directly to a displacement of the photon emission spectrum toward lower energies. As discussed in earlier studies~\cite{Friedlander-II,Johnson}, this implies that gamma-ray constraints derived for standard PBHs are translated to substantially lower masses in higher-dimensional geometries.

Our analysis also shows that the spectral peak exhibits a linear dependence on the black hole temperature,
viz.
$E_\text{peak}\approx 3.96\, T_\text{BH}$.
In the standard scenario we have instead
$E_\text{peak}\approx 5.8\, T_\text{BH}$~\cite{Carr-2}. Thus, not only is the spectrum shifted to lower energies because of their colder temperature, but the overall shape of the curve is also modified.

\begin{figure}[htbp]
    \centering
    \includegraphics[width=0.6\textwidth]{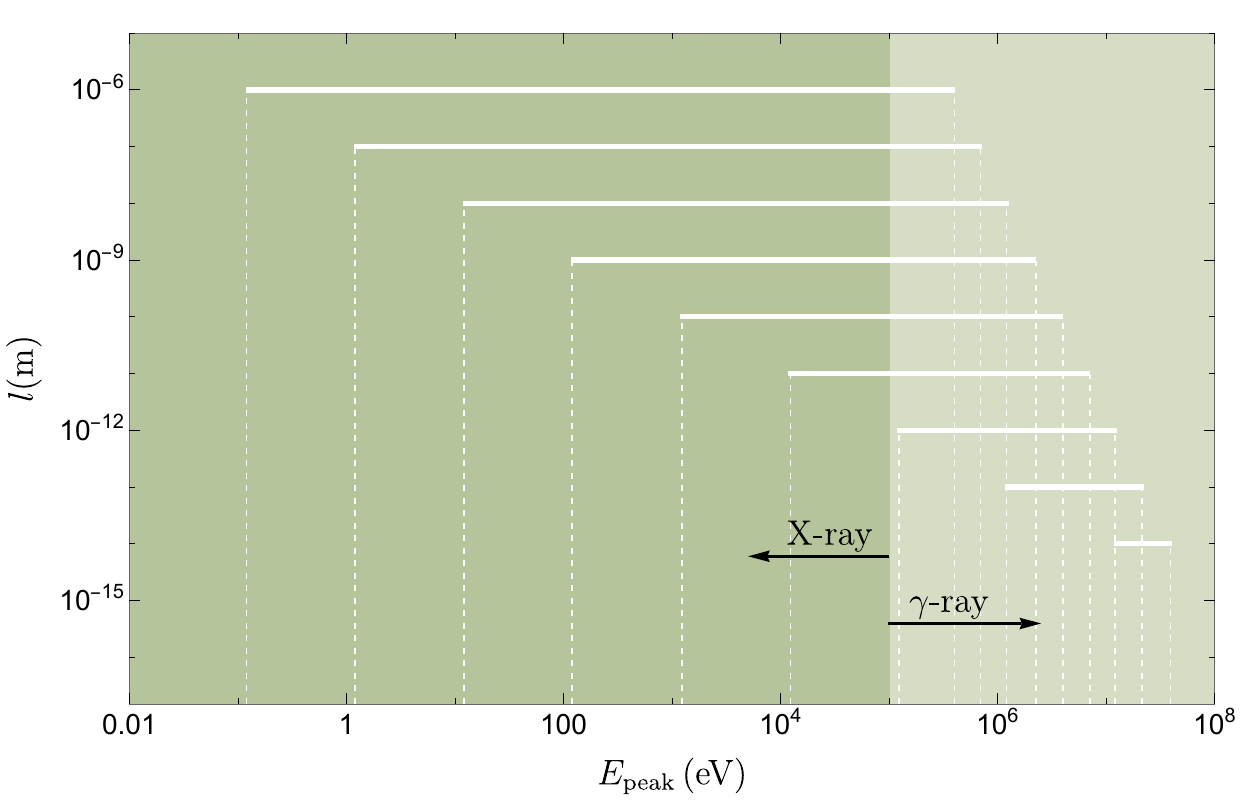}
    \caption{Each line covers the range of $E_\text{peak}$ of five-dimensional PBHs that are DM candidates in RS-II, i.e.~ranging from the lightest mass ($M_\text{BH}\sim3M_*$) to largest possible Schwarzschild--Tangherlini radius ($r_0(M)<l$) for different AdS radius values.}
    \label{fig-Epeak}
\end{figure}

As can be seen in Figure \ref{fig-Epeak},
for larger values of the AdS radius $l$,
there is typically a wide range of masses where the spectral
peak lies in the X-ray band.
This suggests that to
distinguish signatures of PBHs
from direct photon production, it may be more profitable
to focus more on X-ray rather than $\gamma$-ray
observations.
\subsubsection*{Final state radiation}
High-energy charged particles emitted as Hawking quanta
are expected to radiate on-shell photons via
in-vacuo QED effects.
Radiative corrections predict photon emission shortly
after charged particles are produced. In the literature,
this is described
as final state radiation (FSR).
The process is described by the
Dokshitzer--Gribov--Lipatov--Altarelli--Parisi (DGLAP)
splitting functions~\cite{Altarelli-Parisi}.
These represent
the differential probability densities for a charged particle
with energy $E_i$ to emit a photon with energy $E_\gamma$.
To leading order \cite{splitting-functions},
\begin{equation}
    P(i\rightarrow i\gamma)=
    \left\{
    \begin{array}{ll}
        \displaystyle
        \frac{2(1-x)}{x} & \text{(bosons)} \\[2ex]
        \displaystyle
        \frac{1+(1-x)^2}{x} & \text{(fermions)}
    \end{array}
    \right.
\end{equation}
with $x\equiv 2E_\gamma/Q_i$, where $Q_i=2E_i$ is the energy scale. The FSR photon spectra per particle $i$ is given by
\begin{equation}
    f_{\text{FSR}}=\frac{\alpha_\text{EW}}{\pi Q_i}\left(\ln\frac{1-x}{\mu_i^2}-1\right)P(i\rightarrow i\gamma)\,,
\end{equation}
with $\mu_i\equiv m_i/Q_i$. 
This probabilistic photon emission process has to be applied to each emitted particle,
\begin{equation}\label{FSR-em}
\frac{\dd \dot{N}_\gamma ^\text{FSR}}{\dd E_\gamma}=\sum_i\int_{m_i+E_\gamma}^\infty \dd E_i\frac{\dd \dot{N}_i}{\dd E_i}f_\text{FSR}(E_i,m_i,E_\gamma)\,,
\end{equation}
which results in a photon flux with a
broad spectrum of frequencies $E_\gamma$.
This flux is
particularly significant at lower energies.
In our analysis we include $e^\pm$ and $\mu^\pm$ but neglect other fermions and all bosons due to their large masses, which suppress emission. 

For charged particles,
final state radiation is expected to yield
a modest shift to lower energies in the spectrum.
Since it is produced promptly after emission of the Hawking quanta,
it alters the initial conditions for all in-medium interactions.
However, the energy loss in the positron flux is expected to be only a
few percent (see for example Ref.~\cite{FSR-percent}),
even for highly energetic initial states.
Therefore, although we accurately track photons
produced by this mechanism,
for simplicity
we continue to use the undepleted,
initial spectrum
for the positrons themselves.
\subsubsection*{$\bm{e^-e^+}$ inflight annihilation}
Charged particles created via Hawking evaporation are expected to lose energy via in-medium interactions after emission. For positrons with $E_{e^+}\leq 1\,\text{GeV}$ traveling through the ISM, ionization and excitation are the dominant energy-loss mechanisms. This energy loss is quantified via the relativistic Bethe--Bloch formula \cite{Bethe-Bloch}. To leading order, this can be approximated \cite{Beacom}
\begin{equation}\label{Bethe-Bloch}
    \bigg|\frac{\dd E}{\dd x}\bigg|\approx \frac{7.6\times 10^{-26}}{v^2}\left(\frac{n_H}{0.1\,\text{cm}^{-3}}\right)\left(\ln\gamma+6.6\right)\,\frac{\text{MeV}}{\text{cm}}\,,
\end{equation}
where $v$ is the velocity of the incident positron, $\gamma=1/(1-v^2)^{1/2}$ is the Lorentz factor and $n_H$ is the number density of neutral hydrogen in the medium.%
    \footnote{Reducing the complex ISM to pure neutral hydrogen
    is a reasonable approximation as long as the positrons have
    energies $E_{e^+}<1\,\text{GeV}$,
    since ionization and excitation of neutral particles dominates over all other interactions \cite{Jean-2009}.}

For sufficiently small initial kinetic energy, most positrons reach rest
via Bethe--Bloch emission~\eqref{Bethe-Bloch},
and then annihilate or form states with bound or free electrons of the ISM. However, for energies larger than 10 keV, a significant fraction of positrons annihilate inflight before reaching rest.
We calculate the probability of survival for a positron traveling through the ISM with initial energy $E_0$ and final energy $E$ using
\begin{equation}\label{prob-survival}
    P_{E_0\rightarrow E}=\exp \left(-\,n_{e^-}\int^{E_0}_E \frac{\sigma_\text{ann}(E')\dd E'}{\left|\dd E'/\dd x\right|}\right)\,,
\end{equation}
where $\sigma_\text{ann}(E)$ is the annihilation cross-section for a positron of energy $E$ with an electron approximately at rest,%
    \footnote{Even among the hottest ISM environments
    (e.g., supernova remnants), thermal electrons rarely
    reach speeds exceeding a few percent of MeV-scale
    positrons \cite{0007032},
    so this approximation is valid.}
and $n_{e^-}=n_H$ for neutral hydrogen.
We take the tree-level cross-section \cite{ann-cross}
\begin{equation}
    \sigma_{\rm ann}(E_{\rm IA})=\frac{\pi\, r_e^2}{\gamma_{\rm IA}+1}\left[\frac{\gamma_{\rm IA} ^2+4\,\gamma_{\rm IA}+1}{\gamma^2-1}\ln\left(\gamma_{\rm IA}+\sqrt{\gamma_{\rm IA}^2-1}\right)-\frac{\gamma_{\rm IA}+3}{\sqrt{\gamma_{\rm IA}^2-1}}
    \right]
\end{equation}
where $\gamma_{\rm IA}=E_{\rm IA}/m_e$, with $E_{\rm IA}$ being the energy the positron has at the time of annihilation, and $r_e$ is the classical electron radius. 
\\In order to calculate the photon spectrum from the annihilation process, we require the differential cross-section of $e^+e^-\rightarrow 2\gamma$ \cite{ann-cross},
\begin{equation}\label{diff-cross}
    \frac{\dd \sigma_{e^+e^-\rightarrow 2\gamma}}{\dd E_\gamma}=\frac{\pi\, r_e^2}{m_e\,\gamma_{\rm IA}\,v_{\rm IA}}\left[\frac{\left(\gamma_{\rm IA}+1\right)^2+\left(\gamma_{\rm IA}-1\right)^2-2\left(\gamma_{\rm IA}^2-1\right)E_\gamma/m_e}{\left(E_\gamma/m_e\right)^2\left(\gamma_{\rm IA}+1-E_\gamma/m_e\right)^2}
    \right]
\end{equation}
where $E_\gamma$ is the photon energy in the lab frame ($e^-$ at rest) and lies in the range
\begin{equation}\label{energy-bounds}
    \frac{m_e}{2}\left(\gamma_{\rm IA}-1\right)\leq E_\gamma<\frac{m_e}{2}\left(\gamma_{\rm IA}+1\right)\,.
\end{equation}
With \eqref{diff-cross} one can calculate the 
distribution function of the emitted photons \cite{fIA},
\begin{equation}\label{eq-fIA}
    f_{\rm IA}(E_\gamma,E_0)=n_e\int^{E_0}_{E_{\rm min}}\frac{\dd \sigma}{\dd E_{\gamma}}(E_\gamma,E_{\rm IA})\frac{P_{E_0 \rightarrow E_{\rm IA}}}{|\dd E/\dd x|(E_{\rm IA})}\dd E_{\rm IA}\,,
\end{equation}
where $E_{\rm min}$ is determined by \eqref{energy-bounds}. Equation \eqref{eq-fIA} describes a continuous spectrum of photons strongly shifted via the Doppler effect. If we use the notation $P_{E_0\rightarrow m_e} \equiv P(E_0)$, the percentage of the total positron flux that engages in inflight annihilation is $1-P(E_0)$. Therefore,
\begin{equation}\label{IA-em}
    \frac{\dd \dot{N}_{\rm IA}}{\dd E_{\gamma}}=\int_{m_e}^\infty\dd E_0 \frac{\dd \dot{N}_{e^+}(E_0)}{\dd E_{e^+}}(1-P(E_0))\,f_{\rm IA}(E_\gamma,E_0)\,.
\end{equation}
\subsubsection*{Positron thermalization}
From \eqref{prob-survival} we learn that,
after Hawking emission,
most positrons radiate down to low kinetic energy.
When the kinetic energy of the positron becomes comparable
to those in the ambient medium, positrons start to thermalize
(i.e.~their energy distribution becomes Maxwellian \cite{Boehm}).
They are expected to mostly annihilate with ISM electrons or to form
a positronium (Ps) bound state that rapidly decays.%
    \footnote{See \cite{Boehm} for a complete list of possible interactions after thermalization and likelihood in different galactic regions.}

Utilizing data from INTEGRAL/SPI, the fraction of positrons that form positronium before annihilating (positronium fraction, $f_{\text{Ps}}$) in the ISM of the Milky Way was estimated in \cite{positronium}. In the galactic bulge it is measured to be approximately $1.08\pm 0.03$, indicating that effectively all positrons form positronium before annihilation. In the Galactic disk, the positronium fraction is slightly lower but still consistent with almost complete positronium formation, measured at $0.90\pm 0.19$. These findings suggest that, throughout the Milky Way, positrons predominantly disappear via positronium formation and subsequent decay. 

Positronium is formed in a 3:1 ratio\footnote{This ratio is due to statistical spin considerations. Ortho-positronium is a triplet state with spin one (i.e.~three possible spin alignments) and para-positronium is a singlet state with spin zero and has only one alignment.} between ortho-positronium (oPs) and para-positronium (pPs). The former decays into three photons with $E_\gamma<511\,\text{keV}$ and the latter into two photons at exactly 511 keV\footnote{Annihilation into a larger number of photons in both cases is allowed but with negligible corresponding branching ratios \cite{Boehm}.}. 
\\With this in mind, one quickly sees that the number of 511 keV photons produced per positron at rest is $2\times[(1-f_{\text{Ps}})+f_{\text{Ps}}/4]$,
where the first term represents direct annihilation of positrons and the second term photon emission from para-positronium decay. The number of $E_\gamma<511\,\text{keV}$ photons in turn is $3\times (3f_{\text{Ps}}/4)$.

The positron flux that reaches thermalization is $P\times\dd \dot{N}_{e^+}/\dd E_{e^+}$,
and the flux of 511 keV photons from this mechanism is
\begin{equation}\label{flux-511}
    \frac{\dd \dot{N}_{511}}{\dd E_\gamma}=\left[2(1-f_{\text{Ps}})+f_{\text{Ps}}/2\right]\int_{m_e}^{\infty}\dd E_{e^+}\,P(E_{e^+})\frac{\dd \dot{N}_{e^+}}{\dd E_{e^+}}\,f_{511}\,,
\end{equation}
where clearly the probability density distribution in this case is $f_{511}=\delta(E_\gamma-m_e)$. 

On the other hand, ortho-positronium decay results in a continuous spectrum of photon energies first derived by Ore \& Powell \cite{3-insteadof-2}, where the following expression for the normalized probability distribution function was obtained\footnote{More recent calculations were carried out in \cite{Femenia} using non-relativistic quantum field theory. These corrections are important for photons in the UV range. Since the ISM strongly absorbs UV, we cannot employ this range of frequencies for constraints.}:
\begin{equation}
\frac{1}{\Gamma}\frac{\dd\Gamma}{\dd \xi}= \frac{2}{(\pi^2-9)}\left[\frac{\xi(1-\xi)}{(2-\xi)^2}+\frac{2-\xi}{\xi}+\frac{2(1-\xi)^2\ln(1-\xi)}{(2-\xi^3)}+\frac{2(1-\xi)\ln(1-\xi)}{\xi^2}\right]\,,
\end{equation}
with $\xi\equiv E_\gamma/m_e$. The probability density per unit energy is thus
\begin{equation}
   f_{\text{oPs}}(E_\gamma)=\frac{1}{\Gamma_0}\frac{\dd\Gamma}{\dd E}=\frac{1}{m_e} \frac{1}{\Gamma}\frac{\dd\Gamma}{\dd \xi}
\end{equation}
and the integrated flux sourced via oPs decay,
\begin{equation}\label{OPS-em}
    \frac{\dd \dot{N}_\gamma^\text{oPs}}{\dd E_\gamma}= \frac{9f_\text{Ps}}{4}\int_{m_e}^{\infty}\dd E_{e^+}P(E_{e^+})\frac{\dd \dot{N}_{e^+}}{\dd E_{e^+}}\,f_{\text{oPs}}(E_\gamma)\,.
\end{equation}
\subsection{Photon distribution}\label{photon-distribution}
\subsubsection*{Anisotropic component}
If primordial black holes account for the dark matter in the Universe, they are expected to have a specific energy density profile within our galaxy, just as any other dark matter candidate. The emitted electromagnetic flux is expected to have an anisotropic component separable from an isotropic background, with the ratio of the anisotropic and isotropic components depending on the galactic coordinates. One expects intensity variations due to the local density enhancement in the Galaxy.

Due to the proximity of this PBH population we neglect redshifting
between emission and detection. We assume a spherically symmetric
PBH distribution about the centre of the Galaxy and adopt
the Navarro--Frenk--White (NFW)
dark matter density profile \cite{NFW}
\begin{equation}\label{nfw-profile}
    \rho_\text{PBH}(R,f_{\rm PBH})
    =
    \frac{
        f_\text{PBH} \, \rho_\odot
    }{
        \frac{R}{R_\odot}
        \left(
            1
            +
            \frac{R}{R_\odot}
        \right)
    }
\end{equation}
where $R$ is the galactocentric distance,
\begin{equation}
R(s,b,L)=\sqrt{s^2-2s\,R_\odot \cos{b}\cos{L}+R_\odot^2 }
\end{equation}
with $s$ the line-of-sight distance, $(b,L)$ the galactic coordinates\footnote{We use $(b,L)$ instead of the standard notation $(b,l)$ to avoid confusion with the AdS radius $l$.} and $R_\odot$ the distance of the Sun from the galactic centre.
To assign numerical values to the
parameters in \eqref{nfw-profile},
we use the best-fit values obtained by Cautun \textit{et al.} \cite{NFW-parameters}, viz.,
$R_\odot=8.2\,\text{kpc}$
and
$\rho_\odot=0.33\,\text{GeV}/\text{cm}^3$.

In order to obtain the total flux over a region of interest ($s,\Delta\Omega$) we integrate the PBH number density along the line of sight and the sky region we are interested in, i.e.~
\begin{equation}
    \label{phi-gal}
    \Phi_\text{gal}(E_\gamma,\Omega)
    =
    \left(
        \frac{1}{4\pi\Delta\Omega}
    \right)
    \frac{
        \dd\dot{N}^{\text{tot}}_\gamma
    }{
        \dd E_\gamma
    }
    \int_{\rm l.o.s}
    \int_{\Delta\Omega}
    n_{\rm PBH}(s,\Omega)
    \, \dd s
    \, \dd \Omega
    \qquad \text{[units:
    $\text{eV}^{-1} \, \text{cm}^{-2} \, \text{s}^{-1} \, \text{sr}^{-1}$]}
\end{equation}
where we have divided by the solid angle $\Delta\Omega$ in order to obtain the flux per steradian. The brightness associated with this flux follows the standard definition,
\begin{equation}
I(E_\gamma,\Omega)=E_\gamma\,\Phi_\text{gal}\,.
\end{equation}

\subsubsection*{Isotropic component}
In the PBH--DM scenario,
black holes are expected to be distributed throughout the Universe.
As explained above, this will
source a nearly isotropic extragalactic background light.

\begin{figure}[htbp]
    \centering
       \begin{subfigure}[t]{0.48\textwidth}
        \includegraphics[width=\textwidth]{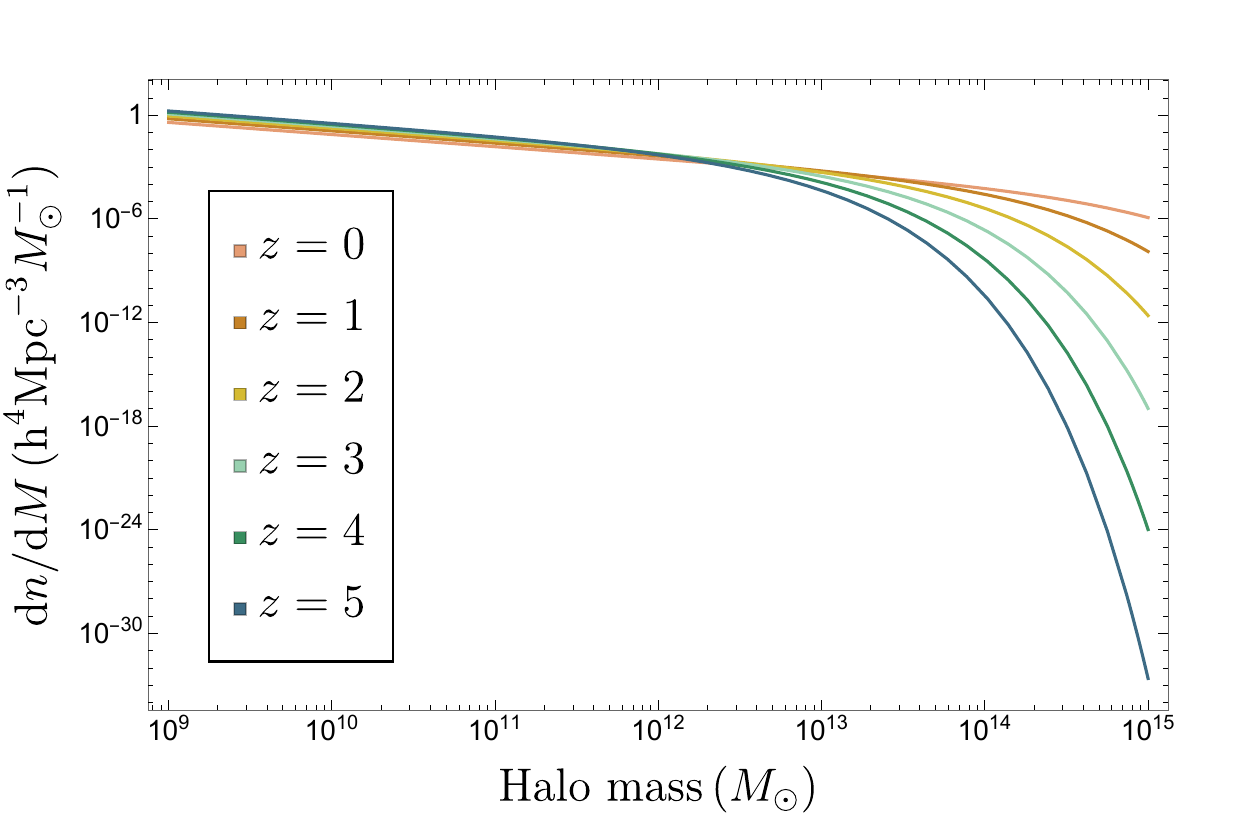}
    \end{subfigure}
    \hfill
    \begin{subfigure}[t]{0.48\textwidth}
        \includegraphics[width=\textwidth]{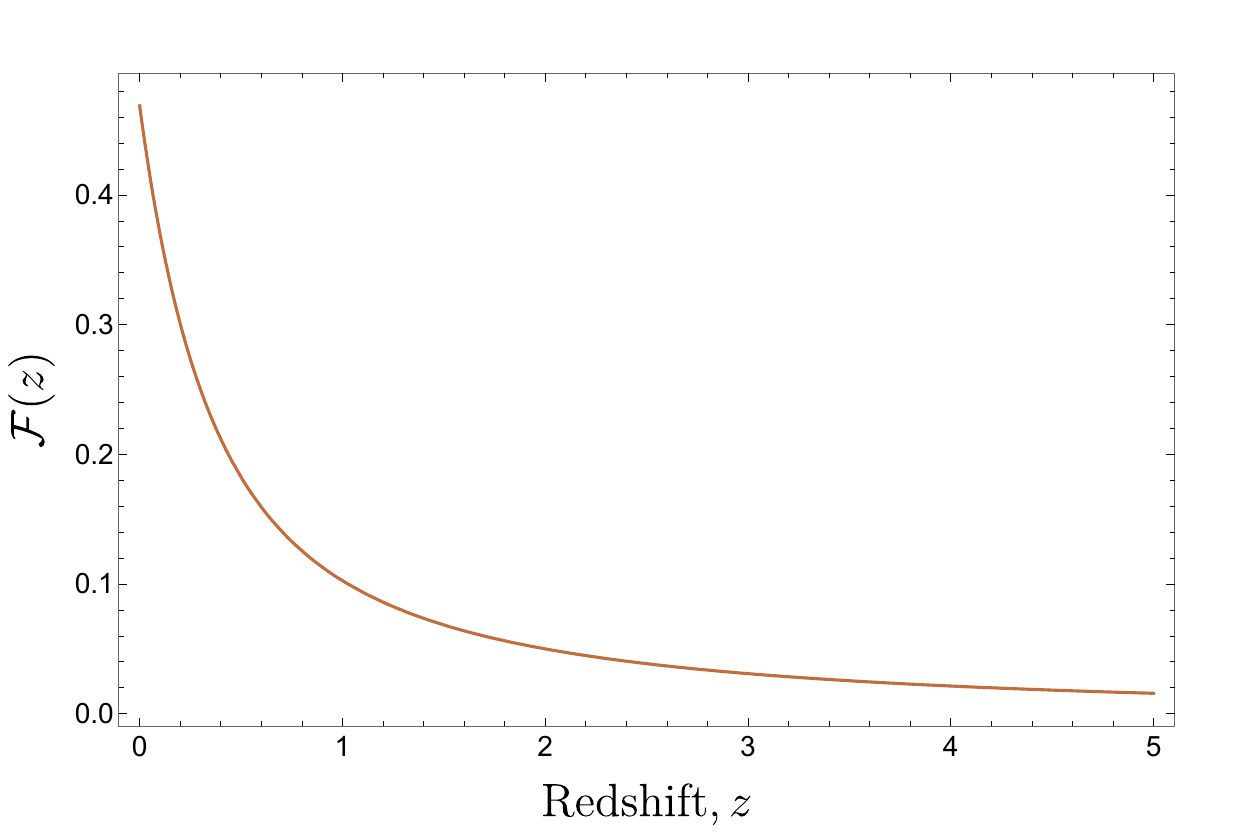}
    \end{subfigure}
    \caption{Comoving number density of haloes per mass interval for different redshift values in the Press-Schechter formalism (left) and the fraction of PBHs located in collapsed haloes for different redshifts (right).}
    \label{HMF-fig}
\end{figure}

Note that the location of the PBHs makes an important difference
when considering the
production mechanisms described in Section~\ref{sec:em-production}.
In our Galaxy,
the interstellar medium has
sufficiently large $n_{H}$ for efficient $e^-e^+$ interactions.
This is not necessarily the case for all extragalactic PBHs.
Positrons must be emitted in environments with a significant particle density, such that $n_{e^-}\geq0.01-0.1\,\rm cm^{-3}$.
This happens mostly in sufficiently massive haloes.
To quantify this, we use the Press--Schechter formalism%
    \footnote{The HMF provided by Tinker et al.~in 2008
    \cite{Tinker} is a better fit to observations but the difference
    is less than an order of magnitude in the mass range we consider \cite{comparison-HMF}.
    We use the Press--Schechter prescription for simplicity.}
\cite{PS-HMF} to define the halo mass function (HMF),
i.e.~the comoving number density of haloes
per mass interval $\dd n/\dd M(M,z)$.
This is shown in Figure \ref{HMF-fig}
as a function of the halo mass at different redshifts.
We see that at high redshift,
the number density of high-mass galaxies drop significantly, as expected. 

We compute the fraction of matter in the Universe contained in
sufficiently massive haloes
by integrating the HMF over a suitable mass range,
\begin{equation}
    \rho_{\rm collapsed}(z)=\int_{M_{\rm min}}^{M_{\rm max}}M\frac{\dd n}{\dd M}(M,z)\,\dd M\,.
\end{equation}
$M_{\text{min}}$
should represent the lowest mass halo that can retain baryons
and cool them sufficiently to collapse into efficient annihilation zones.
We choose the conservative value
$M_{\rm min}\sim 10^{9}M_\odot$~\cite{DM-halos-baryons}.
At the upper end, we introduce a cutoff
$M_{\rm max}\sim 10^{16}M_\odot$.
It follows that
the fraction of PBHs in efficient annihilation zones at redshift $z$ is
\begin{equation}\label{Fofz}
    \mathcal{F}(z)=\frac{(1-f_{\rm b})\,\rho_{\rm collapsed}(z)}{(1-\bar{f}_{\rm b})\,\rho_{\rm m}(z)}\approx \frac{\rho_{\rm collapsed}(z)}{\rho_{\rm m}(z)} \,,
\end{equation}
where $f_{\rm b}$ is the fraction of matter in haloes that is of baryonic nature, $\bar{f}_{\rm b}$ the cosmic mean and $\rho_{\rm m}=\Omega_{\rm m}(z)\rho_{\rm cr}(z)$. For this mass range,
the ratio $f_{\rm b}/\bar{f}_{\rm b}$ is expected to be in the range $0.2$--$1$~\cite{baryon-fraction}. 

We conclude that, when discussing extragalactic PBHs, only
a fraction $\mathcal{F}(z)$
of the total extragalactic positron flux contributes to the photon spectrum via in-medium interactions. In Figure~\ref{HMF-fig} we see that $\mathcal{F}(z)$ is non-negligible for small redshifts. Therefore, the photon flux produced via in-medium interactions in nearby galaxies is particularly relevant when estimating the total extragalactic flux. 

The total flux received from extragalactic sources is then
\begin{equation}\label{EGB-phi}
   \Phi^{\rm EGB}(E_\gamma)=\frac{1}{4\pi}\,n_{\rm PBH,0}\int_0^{z_{\rm max}}\frac{\dd z}{H(z)}\left(\frac{\dd \dot{N}_\gamma^{\rm P,\,FSR}}{\dd E_\gamma}+\mathcal{F}(z)\frac{\dd \dot{N}_\gamma^{\rm IA,\,NR}}{\dd E_\gamma}\right)(E_\gamma(1+z)) \,,
\end{equation}
where
\begin{equation}
H(z)=H_0\sqrt{\Omega_\Lambda+\Omega_{\rm m}(1+z)^3}
\end{equation}
is the Hubble rate of expansion as a function of redshift
with $H_0=67.97\pm0.38\, \rm km/(s\,Mpc)$, $\Omega_\Lambda=1-\Omega_{\rm m}$, $\Omega_{\rm m}=0.307\pm0.005$  \cite{DESI}.

Note, however, that extragalactic PBHs need not be
the only source of isotropic electromagnetic flux. It was noticed by Blanco \& Hooper~\cite{Hooper-2}
that the line-of-sight integral in Equation~\eqref{phi-gal}
deviates less than $10\%$ from its average value
over the high Galactic latitude region $|b|\geq20\degree$.
This is often the area studied in analyses
of the isotropic diffuse background
in order to avoid galactic bulge contamination.
The conclusion is that,
over this region,
a contribution from the galactic halo
would be inseparable from the isotropic extragalactic component.
Its precise contribution depends on the parameter choice,
although in our analysis we see that the enhancement is not more than $\mathcal{O}(1)$.

\begin{figure}[htbp]
    \centering
    \includegraphics[width=1\linewidth]{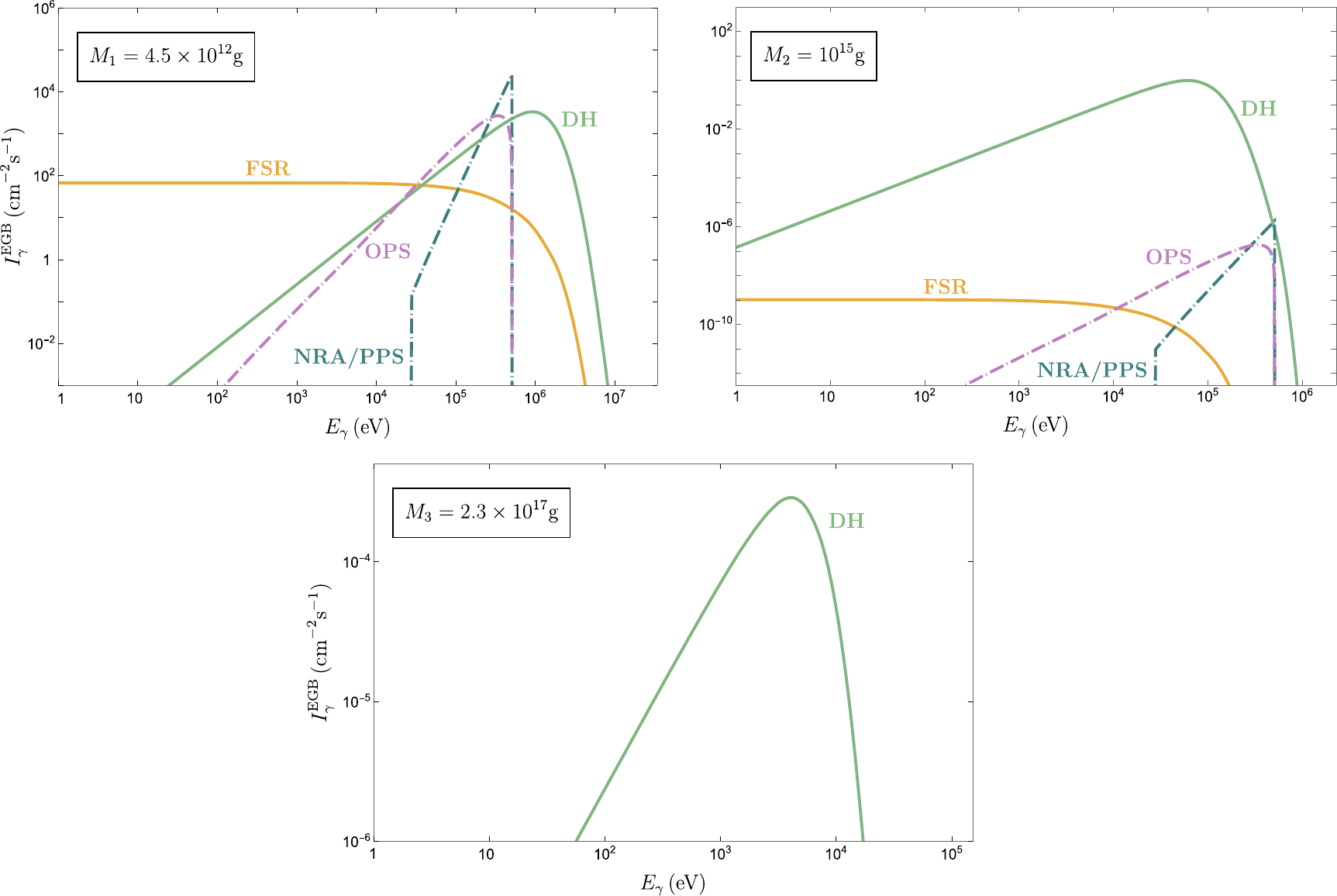}
    \caption{Extragalactic brightness from different photon production mechanisms for three black hole masses (see Figure \ref{emission-allowed} below) and $l=10^{-10}$m. Vacuum processes (FSR and DH) are depicted using solid lines, in-medium emission (NRA/PPS and OPS) using dashed lines. IA is omitted due to its relatively small impact.}
    \label{EGB-plots}
\end{figure}

The total isotropic brightness is then
\begin{equation}
    I^\text{iso}_\gamma(E_\gamma)=E_\gamma (\Phi^\text{EGB}+\Phi^\text{gal}_{|b|\geq20\degree}).
\end{equation}
In Figure \ref{EGB-plots} we can see the flux of different photon emission channels sourced by extragalactic primordial black holes, following Equation \eqref{EGB-phi}. We do not include the contribution from inflight-annihilation in the graphs since it is subdominant in all three cases. The masses in Figure \ref{EGB-plots} are those shown in Figure \ref{emission-allowed}. Clearly, contributions from lepton emission cannot be neglected for hotter black holes. Since $(3+1)$-dimensional PBHs are hotter, this should also have a substantial effect in the conventional scenario. 
%
\subsection{Unresolved electromagnetic signals in the Universe}
\label{sec:unresolved-electromagnetic}
There is unresolved background radiation
permeating the Universe at multiple wavelengths.
This includes both an isotropic component,
and anisotropic fluctuations that presumably
reflect the spatial clustering of unknown sources.
These signals may encode critical information about
exotic phenomena. In this section we consider the possibility that Hawking evaporation from primordial black holes is the primary source of this
radiation.

\par
\begin{figure}[htbp]
    \centering
    \includegraphics[width=0.6\linewidth]{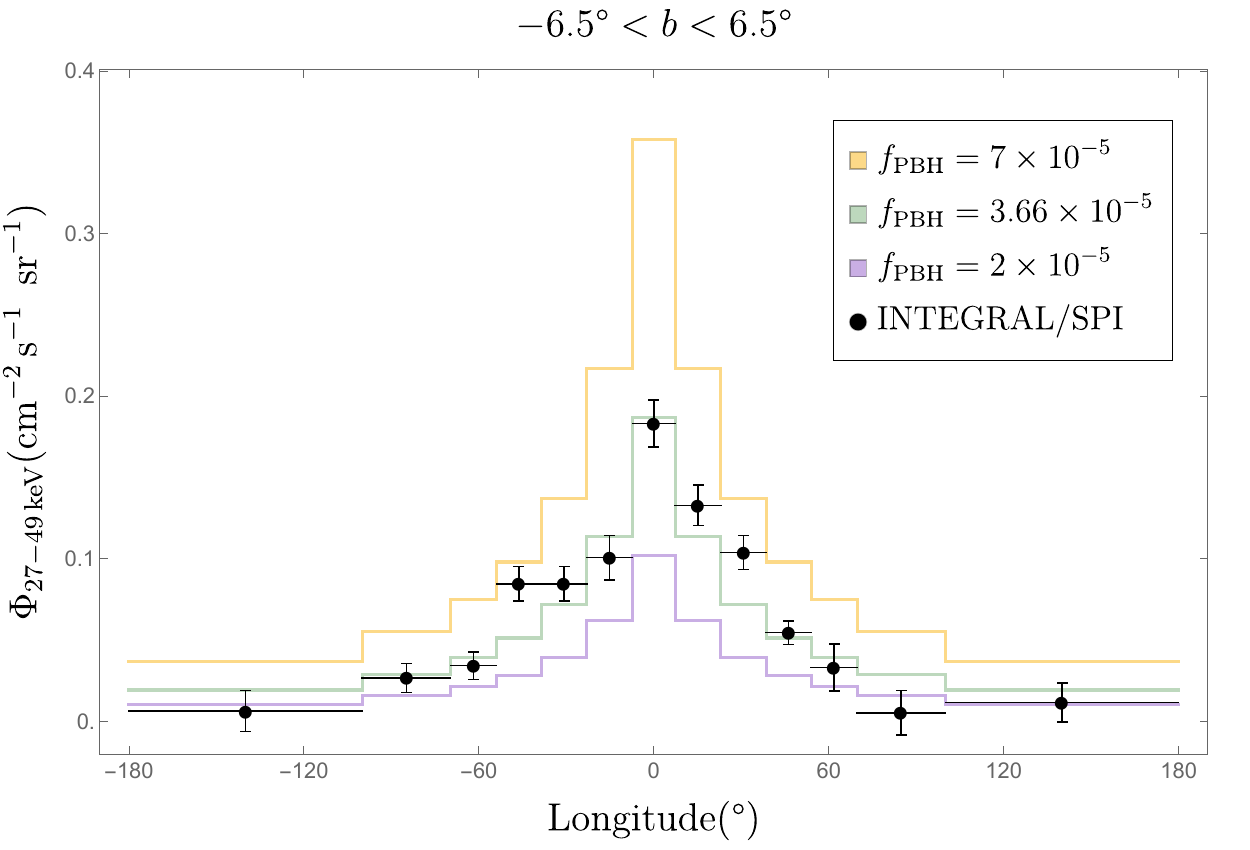}
    \caption{Integrated flux over several sky subregions and energy range 27--49 keV for $M=10^{11}$g and $l=10^{-6}$m for different PBH abundances (solid colored lines), compared with empirical data (black dots). For this parameter choice, the abundance that fulfills $\chi^2_{\rm eff}=4$ is $f_{\rm PBH}=3.66\times 10^{-5}$.}
    \label{fig-GRE-example}
\end{figure}

\noindent\textbf{Datasets.}
We constrain the abundance of PBHs
using a simultaneous fit to
a number of observations.
We use the following measurements of the anisotropic
component.
\begin{itemize}
    \item \textsl{Diffuse emission from the Galactic Ridge.}
    Background-subtracted measurements
    in this region were made the SPI spectrometer
    on board the INTEGRAL observatory~\cite{data-GRE}.

    We fit a PBH population following an NFW distribution as
    the main source.%
        \footnote{This analysis was carried out for $(3+1)$-dimensional PBHs by Laha et al.~in \cite{Laha-GRE}.}
    To match SPI measurements of both the spatial morphology
    and spectral shape associated with the emission,
    we integrate over four energy bins
    ($27$--$49$~keV,
    $49$--$90$~keV,
    $100$--$200$~keV, and
    $200$--$600$~keV),
    and 13 $\Delta L$ bins covering the
    region $-6.5\degree < b < 6.5\degree$. See
    Figure~\ref{fig-GRE-example}.
    In this figure one can clearly see the strong
    sensitivity to the abundance $f_{\text{PBH}}$.

    \item \textsl{Chandra}. We use the background-subtracted
    spectrum reported in Ref.~\cite{sterile-neutrino}
    covering the range $3$--$5.5$~keV at latitudes $|b| \geq 10\degree$.

    \item \textsl{e-ROSITA and XMM-Newton}. We use data
    reported by e-ROSITA
    in the
    $0.2$--$10$~keV range,
    and by XMM-Newton
    in the
    $0.15$--$12$~keV range.
    The strictest constraints come from
    measurements in the sky area
    delimited by the third ring contour
    (see Ref.~\cite{XMM-Newton-data}.)
    For comparison
    to the PBH abundance, we employ the spectra expressed in experiment-independent units calculated in Ref.~\cite{extract-eRos-XMM}.

    Unfortunately,
    bounds on the PBH abundance obtained from the eROSITA/XMM-Newton are not competitive with those arising from Chandra or the Galactic Ridge signals anywhere in our parameter range.
\end{itemize}
We also use the following measurements of the isotropic component:
a Chandra
measurement at $0.5$--$3$~keV~\cite{cappelluti};
a HEAO 1 measurement at $3$--$500$~keV~\cite{223};
a COMPTEL measurement at $0.8$--$30$~MeV~\cite{224};
an EGRET measurement at $30$--$200$~MeV~\cite{225};
and a Fermi-LAT measurement at $200$--$10^4$~MeV~\cite{226}.

\par\noindent\textbf{Model fitting.}
Given a value of $f_{\text{PBH}}$
and the PBH population models described above,
we obtain a theoretical prediction for the flux
observed in each of these measurements.
We report an upper limit
on $f_{\text{PBH}}$
corresponding to the
$2\sigma$
($95.45\%$) confidence limit
for the one-sided goodness-of-fit
statistic
\begin{equation}
    \label{eq:unresolved-chisq}
    \chi^2_{\rm eff}=\sum_i\left(\frac{\text{Max}[\mathsf{X}^{\rm th}_i(f_{\rm PBH})-\mathsf{X}^{\rm obs}_i,0]}{2\,\sigma_i}\right) ,
\end{equation}
based on a global fit.
Here, $\mathsf{X}_i$
represent the $i^{\text{th}}$
observable,
$\sigma_i$ is the corresponding uncertainty,
``th'' denotes the prediction of our population model,
and ``obs'' denotes the measured value.
For the purposes of obtaining an estimate
we ignore covariances between measurements.

Because of the one-sided
fit,
we
drop negative bins where the model underpredicts
the measured flux.
This is because emission from PBHs
does not need to explain the measured data in
all wavebands.
Although the PBH prediction should not significantly
exceed the measured
value in any bin,
any deficit can be attributed to other unresolved sources.

\par
\begin{figure}[htbp]
    \centering
    \includegraphics[width=0.7\linewidth]{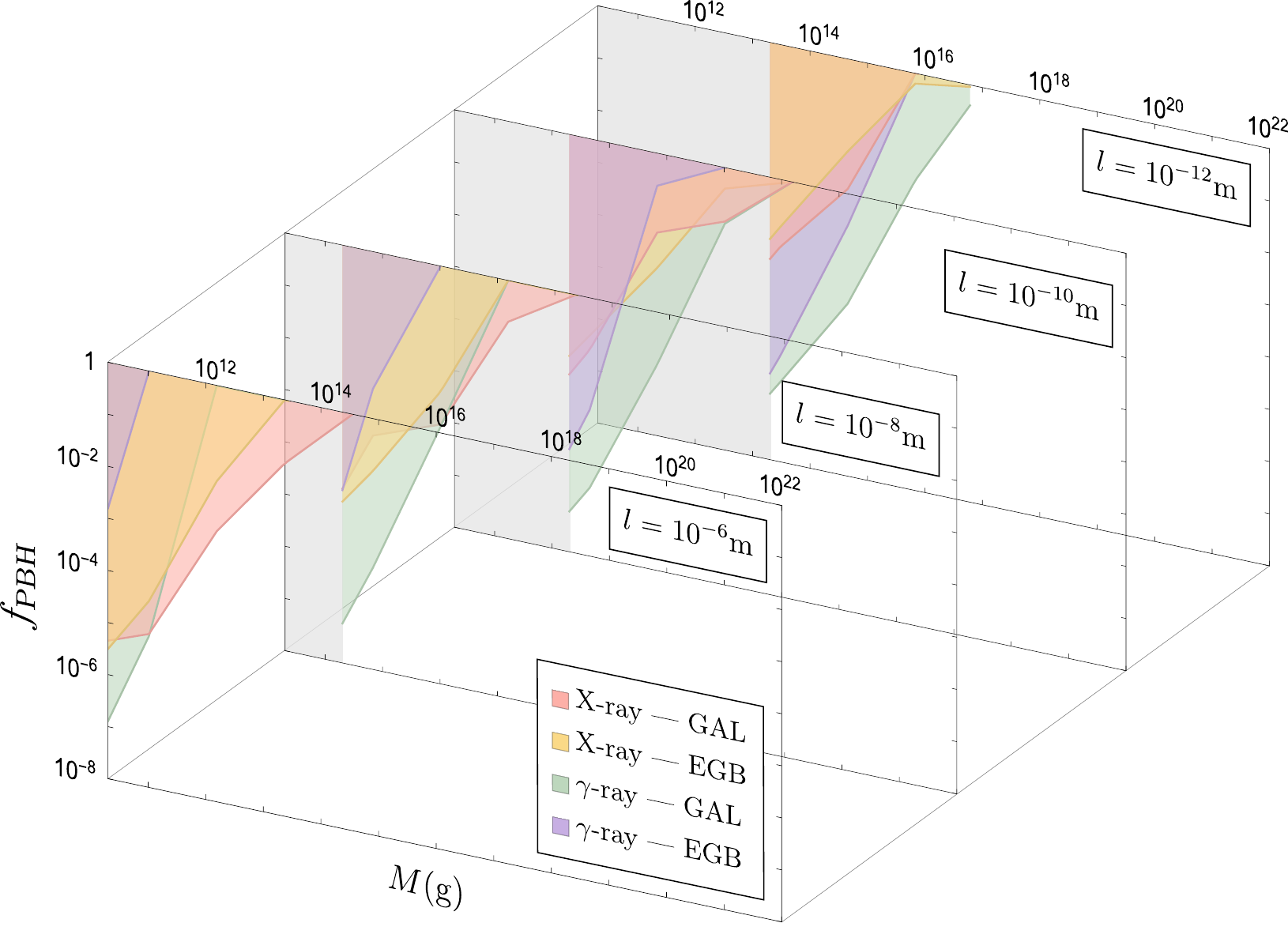}
    \caption{Observational bounds on PBH abundance for a monochromatic population for different AdS radii (from front to back: $l=10^{-6},10^{-8},10^{-10}$ and $10^{-12}$m). Shaded regions are excluded by observations: purple by $\gamma$-ray detection of extragalactic sources, green by $\gamma$-radiation of galactic signals, 
   yellow corresponds to X-ray radiation from extragalactic sources and pink to X-ray radiation from galactic  sources. The grey-shaded regions cover the range of PBHs that have fully evaporated today or that would be rapidly evaporating today, $M<3M_*$.}
    \label{fig-3d-rad}
\end{figure}

\noindent\textbf{Results.}
Our results are shown in
Figure \ref{fig-3d-rad}, where we have plotted
bounds on $f_{\text{PBH}}$
for $l \in \{ 10^{-6} \, \text{m}, 10^{-8} \, \text{m},
10^{-10} \, \text{m}, 10^{-12} \, \text{m} \}$,
here plotted from front to back.%
    \footnote{For the case $l=10^{-12}\,\text{m}$,
    the envelopes end abruptly at $M=6\times 10^{16}$g.
    At this point
    the small black hole limit approximation is no longer valid, i.e.~the Schwarzschild--Tangherlini description breaks down as discussed in Section \ref{section-PBH-BW}.}

The strong dependence on the AdS radius is clearly
visible.
In particular, the dark matter window $f_{\text{PBH}}=1$ is wider
for larger AdS radi $l$ (i.e.~colder black holes).
Also, the relative constraining power of the different observational
data notably shifts. For a fixed mass, a lower temperature results in
emission of radiation at lower frequencies. At
large AdS radius, in contrast with the standard
$(3+1)$-dimensional scenario and small AdS radius,
radiation in the $\gamma$-ray band
has a limited constraining power.
Note that X-ray bounds, even if subdominant,
still usefully constrain the abundance of hotter black holes.
This is because particle production is favoured,
which results in FSR, IA and NRA photon emission
in the X-ray range, as shown in Figure \ref{EGB-plots}. 

\par\noindent\textbf{511 keV excess.}
Since the 1970s,
balloon and satellite experiments have consistently confirmed an excess of $511\,\text{keV}$
photons in the Milky Way~\cite{balloon-1972}. This $511\,\text{keV}$
emission features both disk and bulge components, with the bulge being
exceptionally narrow and bright. The most detailed observations have been
provided by INTEGRAL/SPI \cite{integral-spi}. One of the biggest puzzles
is why the majority of the $511\,\text{keV}$ emission is concentrated in
the galactic bulge, rather than being evenly spread throughout the disk,
as would be expected if supernovae were the dominant source. A variety of
astrophysical sources have been studied \cite{astro-511},
but all candidates face significant challenges
in explaining the observed distribution and intensity of the
signal \cite{review-511}.

Dark matter models may naturally explain the dominance
of emission from the bulge,
because they typically predict a cusp in the dark matter density
near the Galactic centre.
The most studied
models are those involving
light particles, such as MeV-scale
WIMPs or axion-like particles~\cite{lightDM-511}.
These may decay into positrons,
followed by annihilation in the galactic medium
and decay into photons.
Primordial black holes were have
also been considered as a possible origin~\cite{okele}.
These results have been adapted to newer developments in the
field~\cite{ift,hooper511}.

To explore the possibility that braneworld black holes may explain the
$511\,\text{keV}$
excess,
we apply
the following standard procedure.
We integrate the total flux over galactic latitudes in the range $-10.75\degree<b<10.25\degree$,
in an energy bin with width corresponding
to the uncertainty of the INTEGRAL/SPI detectors
at $E_\gamma=511$ keV.
We consider bins of angular size $\Delta L=12\degree$ and compare
our theoretical prediction
to the observed flux in each bin \cite{1906.00498}.

Let us highlight that by introducing the survival
probability~\eqref{prob-survival} in our calculation
(including the IA photon spectrum and integrating over the
contributions from all channels),
we do not have to impose a cutoff on
the initial positron injection,
as discussed in Refs.~\cite{Friedlander-I,graham}.
We obtain results comparable to those 
for the UV-complete spectra
in \cite{Friedlander-I,graham}, which suggests
that the condition $E_{e^+}<1$ MeV
used by these authors underestimates the total $511\,\text{keV}$ flux. The survival probability is still relatively high for such energies.
The probability of reaching thermalization decreases as the energy
increases, but even for energies $E_{e^+}\sim 1$ GeV,
$P_{E_{e^+}\rightarrow m_e}\approx 0.64$. This still leads to a significant contribution to the integral in Equation \eqref{flux-511}.  
%
%
%
%
\section{Low-energy electrons and positrons}\label{section-voyager}
Direct detection
of electrons or positrons
by spacecraft such as
Voyager also places constraints on the abundance
of evaporating primordial black holes.
In fact, this produces one of the most stringent limits on
$f_{\rm PBH}$ in the $(3+1)$-dimensions~\cite{voy-pbh-first-paper}.

The Voyager spacecraft were designed to detect sub-GeV $e^\pm$,
among other charged cosmic rays.
This is ideal for studying PBHs,
since dark matter candidates are expected to produce
leptons in this energy range.
However, direct detection is severely hindered by the very low
probability of survival of positrons,
and the expected energy loss and
subsequent thermalization of electrons.
As a result, we expect $e^\pm$ emitted only in nearby regions,
perhaps up to a few kiloparsecs away from the spacecraft,
to reach the detectors.

Measurements within the solar system are complicated
by the effect of the solar wind on charged cosmic rays.
However,
both Voyager
spacecraft have long crossed the heliopause;%
    \footnote{The \emph{heliopause} is the outermost boundary
    of the heliosphere, the region dominated by solar wind and the
    Sun’s magnetic field.}
Voyager 1 did so in 2012, and Voyager 2 in 2018.
They
have since been gathering more reliable data~\cite{voy-data}.
We calculate the local flux at the
position of the solar system,
and estimate limits on the PBH abundance.
We use
the Parker transport equation to model
steady injection and energy loss.
The total local flux is approximately given by
\begin{equation}\label{voy-flux}
   \Phi^{\odot}_{e^\pm}(E)\approx\int_E^\infty \frac{P_{E'\rightarrow E}}{ |\dd E/\dd x|(E')}Q(E',R_{\odot})\dd E'=\left(\frac{1}{4\pi}\right)\,n_{\rm PBH}(R_\odot)\int_E^\infty\frac{P_{E'\rightarrow E}}{ |\dd E/\dd x|(E')}\frac{\dd \dot{N}_{e^\pm}}{\dd E'}\dd E'\,,
\end{equation}
The label `$\odot$' denotes evaluation at the position of the
solar system, not the solar value,
and (as in Equation~\eqref{nfw-profile})
$R_{\odot}$ is the distance of the solar system
from the galactic centre.
$Q(E,R_{\odot})$ is the local source term,
i.e.~the number of particles per unit energy,
per unit time, per unit volume.
In \eqref{voy-flux} we have used again the
survival probability~\eqref{prob-survival}
and the Bethe--Bloch formula \eqref{Bethe-Bloch}.
In both cases, the estimated number density of neutral hydrogen
at the
position of the solar system
is taken to be
$n_H\approx 0.12\,\text{cm}^{-3}$~\cite{nH-helio}. 
%
%
\subsection{Detection at the heliopause}
We report a
$2\sigma$ upper limit
on $f_{\text{PBH}}$
($>$95.45\% confidence level)
using the same technique described
in Section~\ref{sec:unresolved-electromagnetic};
see Equation~\eqref{eq:unresolved-chisq}.

\begin{figure}[htbp]
    \centering
\includegraphics[width=0.65\linewidth]{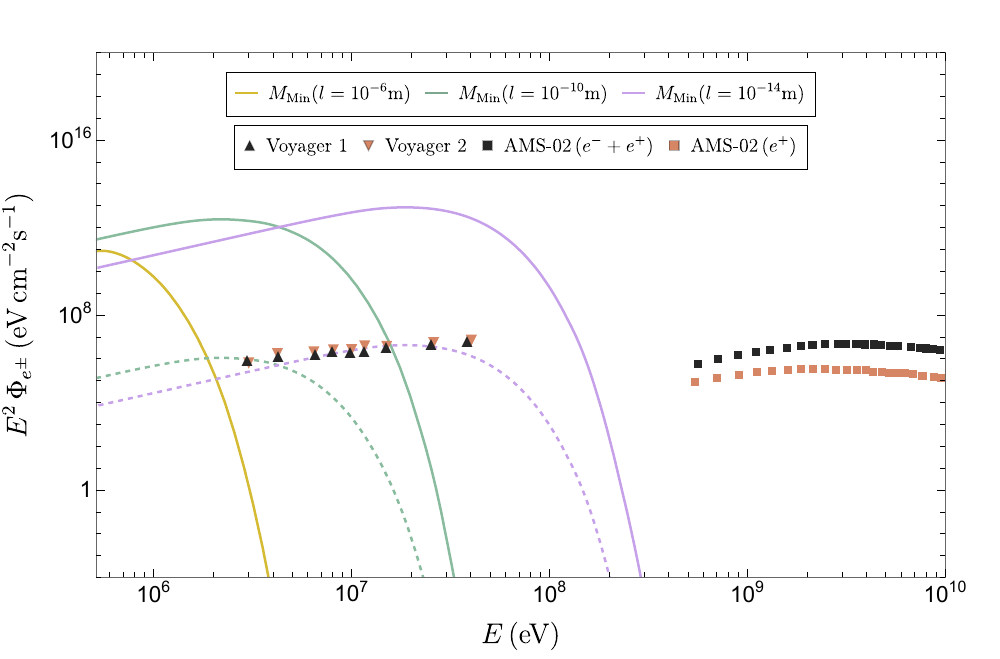}
    \caption{Voyager data and the spectra of $e^\pm$ of the lightest PBH--DM candidates for different values of AdS radius  (after propagation through the galactic medium with no reacceleration). Solid lines are for a PBH population with $f_{\rm PBH}=1$ and the dashed lines for $f_{\rm PBH}$ values such that the flux agrees with observations, with ($l(\text{m}),f_{\rm PBH}$) approximately being $(10^{-10}, 1.7\times 10^{-6})$ and $(10^{-14}, 5.3\times 10^{-7})$.}
    \label{fig-voy}
\end{figure}

The result is plotted in
Figure \ref{fig-voy}. For large AdS radii,
the emitted $e^\pm$ are too cold to be constrained by Voyager.
This differs significantly
from the $(3+1)$-dimensional case \cite{voy-pbh-first-paper},
where the behaviour is similar to the case
$l=10^{-14}\,\text{m}$,
with $f_{\rm PBH}\approx  6.6\times 10^{-7}$
for the lightest steadily evaporating PBH.

Note that diffuse reacceleration of leptons by the
astrophysical galactic background can have a significant
impact on the calculation of upper bounds
for $f_{\rm PBH}$.
This effect is particularly important
for the emission of high-energy leptons in the range $0.1$--$1$ GeV.%
    \footnote{See Figure 1 in Ref.~\cite{voy-pbh-first-paper},
    or Figure 2 in Ref.~\cite{ift}.}
In our case, this corresponds to the tail of the spectrum
$E>E_{\rm peak}$
for $M\sim M_{\rm min}$ at small AdS radii.
It is not relevant
for colder black holes.

For the former, we expect the emission of energies in the tail to be
enhanced (resulting in a broadened peak),
potentially reaching energies constrained by
AMS-02 detections~\cite{ams-data}.
For our parameter space, these bounds are expected to be
comparable or weaker to those from the Voyager spacecraft.
Meanwhile, for the case of colder black holes,
reacceleration of emitted leptons is subdominant,
as energy-loss mechanisms dominate the particle's evolution.
The influence of reacceleration is negligible unless the
turbulence is unusually strong (i.e.~high Alfv\'en speed environments),
which is not the case at the heliopause.
For this reason, braneworld primordial black holes are
expected to avoid useful bounds from AMS-02 positron data,
unlike the $(3+1)$-dimensional
scenario \cite{voy-pbh-first-paper,voy-pbh-second-paper}.
\begin{figure}[ht]
    \centering
    \includegraphics[width=0.6\linewidth]{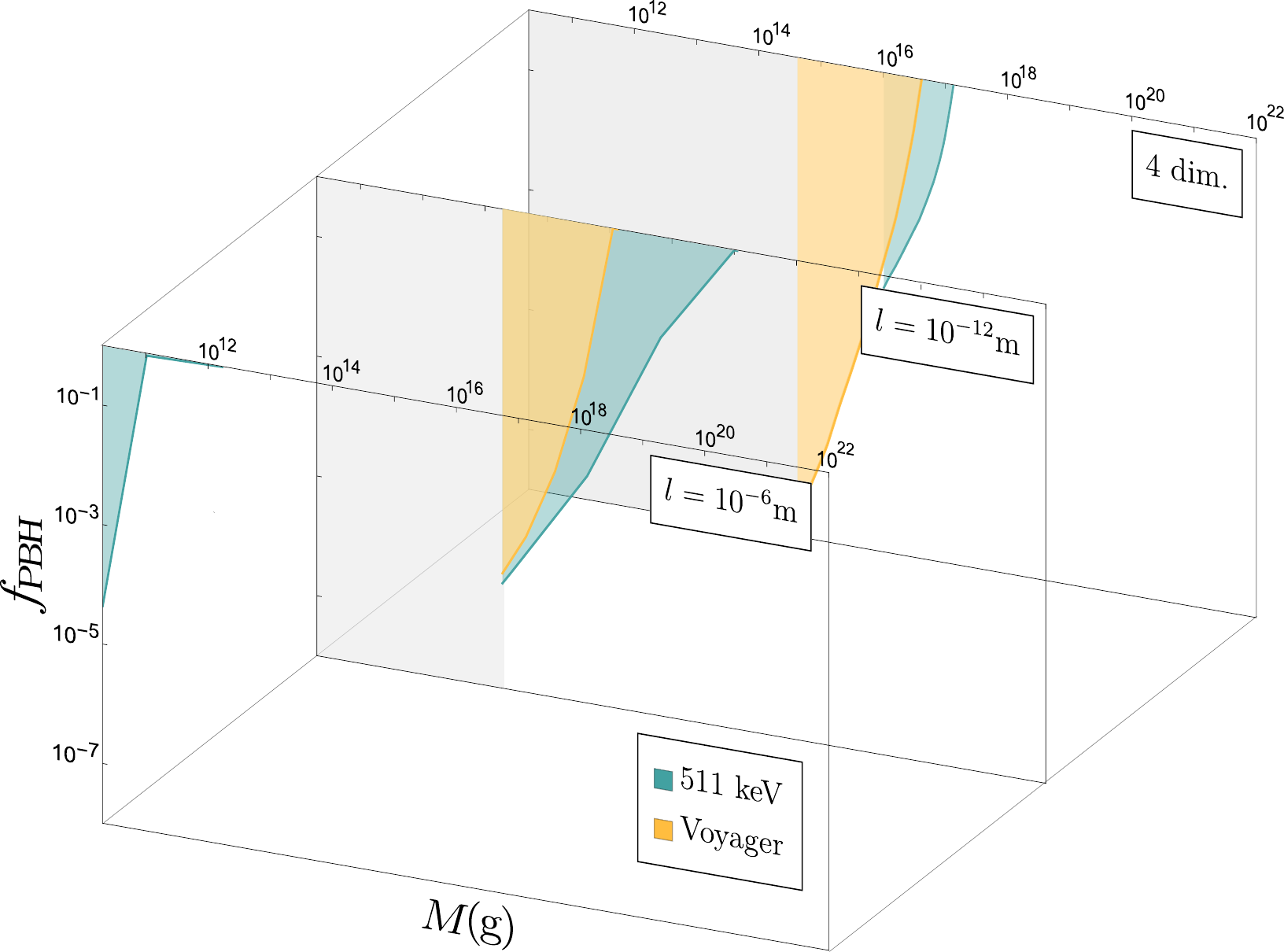}
    \caption{Observational bounds on PBH abundance arising from $e^+e^-$ injection into the medium. We compare the cases $l=10^{-6}$m,
    $10^{-12}$m
    with the bounds obtained for the conventional scenario (511 keV line by DeRocco and Graham \cite{graham} and Voyager by Boudaud and Cirelli \cite{voy-pbh-first-paper}) As already shown in Figure \ref{fig-voy}, for $l=10^{-6}$m, PBHs  are unconstrained by direct $e^+e^-$ detection.}
    \label{fig:enter-label}
\end{figure}
%
%
\section{Impact on the cosmic microwave background}\label{section-CMB}
Predictions for energy injection into their environment
from
evaporating primordial black holes
can yield useful bounds on their abundance.
Electromagnetic energy injection
notably affects the evolution of the
free electron fraction $x_e\equiv n_e/n_H$,
the temperature of the intergalactic medium,
or even the CMB power spectrum. The rate at which energy is injected is
\begin{equation}\label{dEdVdt}
    \frac{\dd^2E}{\dd V\dd t}\Bigg|_{\rm inj}^i=n_{\rm PBH,0}\,(1+z)^3\int E_i\frac{\dd\dot{N}}{\dd E_i}\dd E_i .
\end{equation}
Here $i$ labels the particle
species (photons, electrons, positrons),
and we integrate over the entire spectrum. 

The energy
\eqref{dEdVdt}
is deposited in its environment in three different ways.%
    \footnote{See \cite{CMB-1} for a detailed description of these channels.}
These are: (a) ionization (channel $i$),
(b) heating of the medium (channel $h$),
and (c) excitation of the Lyman-$\alpha$ transition (channel $\alpha$).
The relative importance of these channels
is accounted for by
dimensionless energy deposition
functions $f_c(z)$, 
\begin{equation}\label{dep-energy}
    \frac{\dd^2 E}{\dd V\dd t}\Bigg|_{\rm dep,c}(z)=f_c(z)\frac{\dd^2E}{\dd V\dd t}\Bigg|_{\rm inj}(z)\,,
\end{equation}
where $c=i,h,\alpha$. The $f_c(z)$
quantify how the energy deposition is
split between the channels.
They can be computed by \cite{CMB-2}
\begin{equation}\label{def-fc}
    f_c(z)
    =
    \frac{
        \displaystyle
        \int_z^\infty \dd \ln(1+z')
        \,
        H^{-1}(z')
        \sum_i
            \int\dd E
            \,
            T_c^{i}(z',z,E)
            \,
            E
            \,
            \left.
                \frac{
                    \dd \dot{N}(E,t(z))
                }{
                    \dd E
                }
            \right|_{\rm inj}^{i}
    }{
        \displaystyle
        H^{-1}(z)
        \int\dd E
        \,
        E
        \,
        \left.
            \frac{
                \dd\dot{N}(E,t(z))
            }{
                \dd E
            }
        \right|^{\rm tot}_{\rm inj}
    }
    \,
    ,
\end{equation}
where $T_c^i$ are transfer functions for each particle $i$ and channel $c$. We employ the transfer functions computed in \cite{cmb-transferfunction1,cmb-transferfunction2}. 
%
%
\subsection{Statistical analysis}
To calculate an upper bound on
$f_{\text{PBH}}$
from the CMB
we follow the procedure of Ref.~\cite{CMB-1},
later refined in Ref.~\cite{CMB-2}.
In the context of higher dimensions, a similar analysis was carried
out by Friedlander {\etal}~\cite{Friedlander-I} in the LED scenario.
In each case, modifications to the
\textsc{CLASS} Boltzmann code
(Cosmic Linear Anisotropy Solving System~\cite{CLASS})
were implemented to model energy injection by PBHs.

For this analysis, we adapt the
\textsc{ExoClass}\_\textsc{led}
and \textsc{DarkAges}\_\textsc{led}
modules of \textsc{CosmoLed}~\cite{Friedlander-I}
to the Randall--Sundrum framework.
We consider a range of values for the black hole mass $M$
and AdS radius $l$,
and find the maximum abundance that keeps the CMB power
spectrum within 95\% C.L relative to the Planck 2018 data.

\begin{figure}[ht]
    \centering
\includegraphics[width=0.6\linewidth]{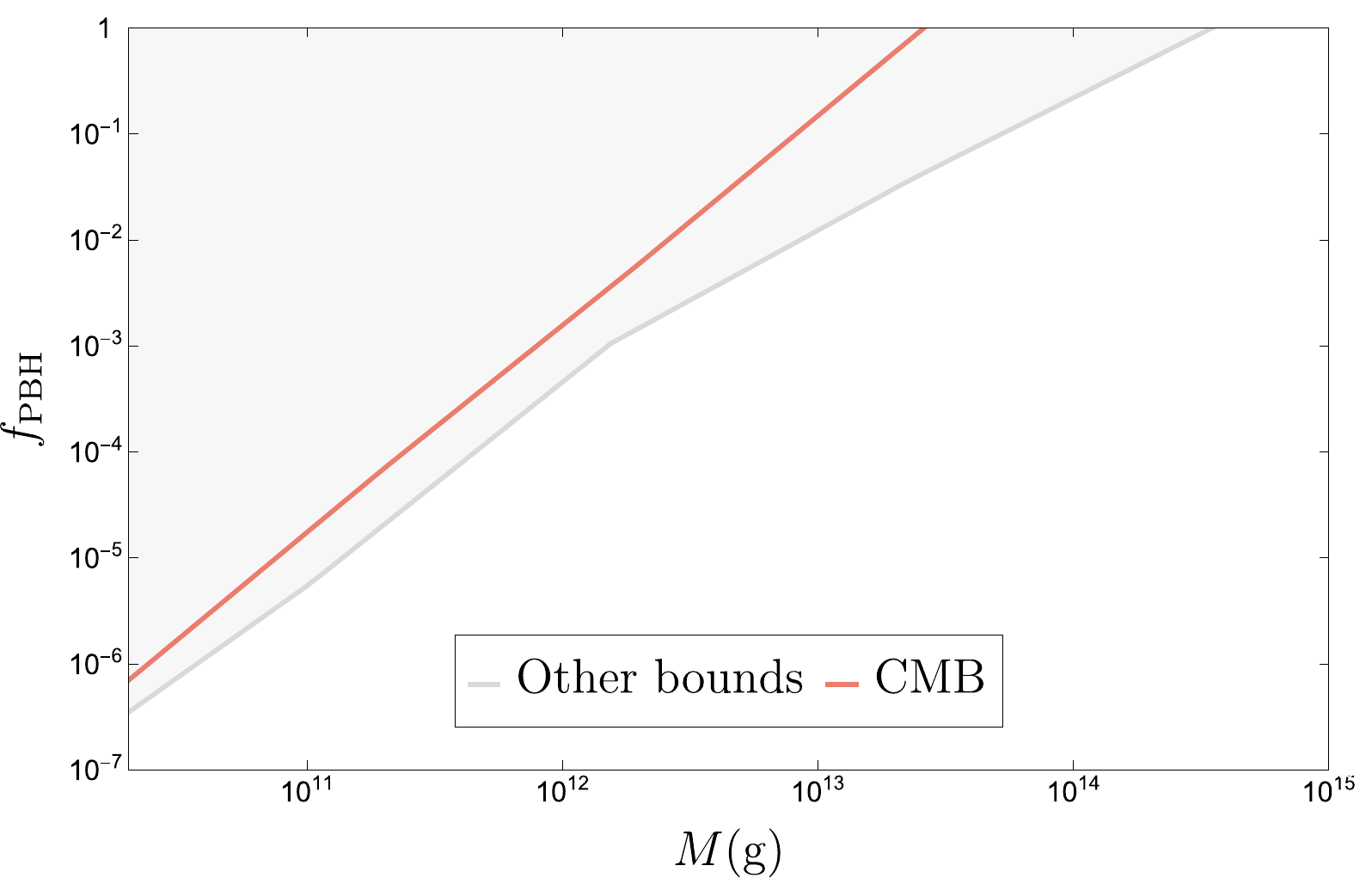}
    \caption{PBH abundance bounds for AdS radius $l=10^{-6}$m. In grey,
    the region excluded by evaporation constraints. In red, conservative bound imposed by our statistical analysis of the CMB power spectra.}
    \label{cmb-line-l6}
\end{figure}

We keep the parameters of the background
$\Lambda$CDM cosmology fixed.
In principle,
one should carry out a Markov Chain Monte
Carlo analysis, as done in Refs.~\cite{CMB-2,Friedlander-I},
which would allow changes in the background cosmology to be
tensioned against the PBH energy deposition.
However, we will find that the CMB
bound does not dominate anywhere in our
$(M, l)$ parameter space
(see Figure~\ref{cmb-line-l6}).
Our analysis can therefore be regarded
as a consistency check
rather than a primary constraint,
for which purpose the simpler approach
is adequate.
It is conservative in the sense that
larger $f_{\text{PBH}}$ may be allowed,
possibly at the expense of tensions with other
cosmological datasets.
For example,
if we compare the bounds
on $f_{\text{PBH}}$
with and without a Monte Carlo analysis in
Ref.~\cite{CMB-2} and
Ref.~\cite{Friedlander-I}, the upper limit
relaxes by an order of magnitude.
We expect similar behaviour in our framework.

\begin{figure}[ht]
    \centering
    \begin{subfigure}[t]{0.48\textwidth}
        \includegraphics[width=\textwidth]{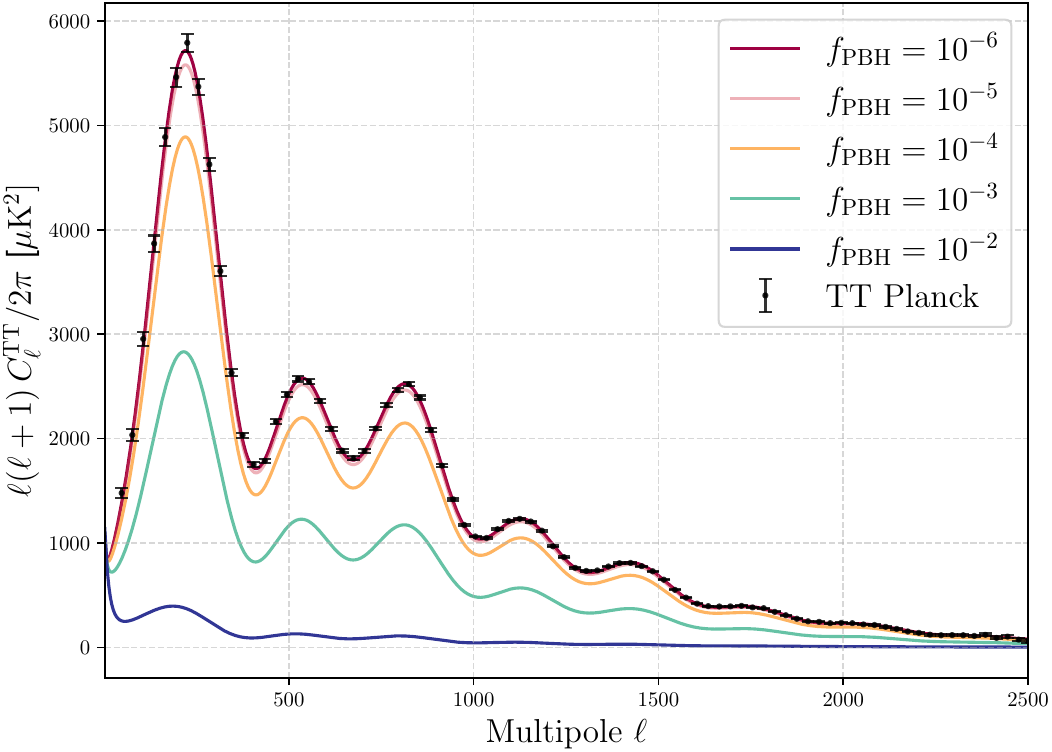}
    \end{subfigure}
    \hfill
    \begin{subfigure}[t]{0.48\textwidth}
        \includegraphics[width=\textwidth]{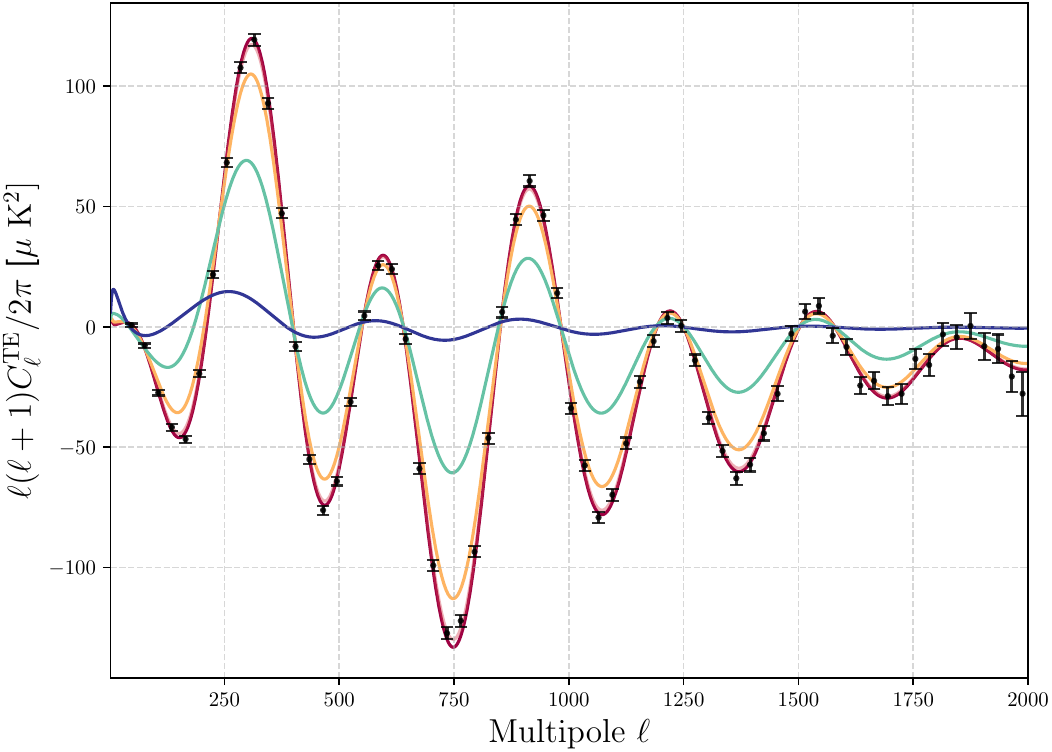}
    \end{subfigure}
    \\
    \vspace{0.5cm}
    \begin{subfigure}
     {0.48\textwidth}
        \includegraphics[width=\textwidth]{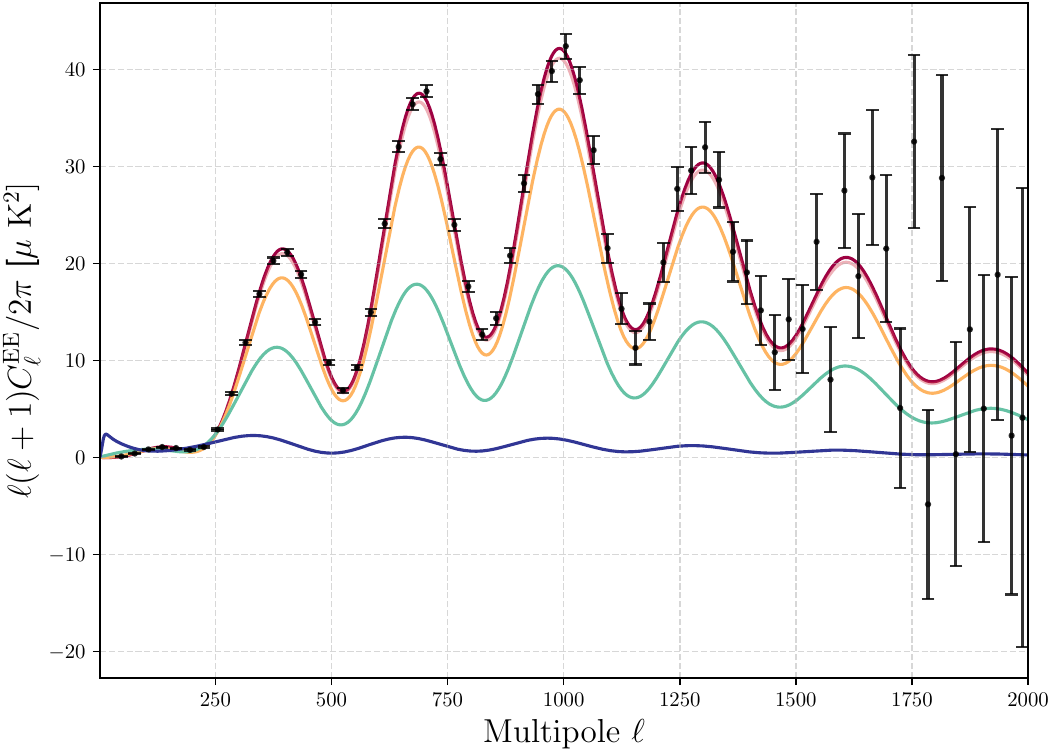}
    \end{subfigure}
    \caption{Example of spectra employed for the $\chi^2$ statistical analysis with Planck 2018 data. The coloured curves are the predicted TT-/TE-/EE-spectra if $\Lambda$CDM is combined with energy injection from a monochromatic PBH population with $(M_{\rm min}(l),l)=(2\times 10^{10}\,\text{g},10^{-6}\,\text{m})$ for different abundances.}
    \label{CMB-fig-1}
\end{figure}

Figure~\ref{CMB-fig-1}
plots
the modified temperature,
polarization and cross-correlation spectra
for a range of values of
$f_{\text{PBH}}$.
In these plots,
damping of anisotropies as the PBH abundance increases
is clearly visible.
%
%
\section{KM3-230213A event}
Recently there has been renewed interest in primordial black holes
within higher-dimensional frameworks~\cite{higher-d-km3,higher-d-km3-2},
because these are a possible explanation for the
KM3NeT event~\cite{km3-detection}.
This is a detection of the most energetic neutrino to date.
The absence of multimessenger counterparts at the time of
detection~\cite{gamma-1,gamma-2},
combined with the lack of similar high-energy events in
IceCube~\cite{km3-icecube,km3-icecube-2},
poses significant challenges for conventional interpretations.
Proposed exotic scenarios include higher-dimensional black holes
with sterile neutrinos \cite{higher-d-km3},
quasi-extremal PBHs \cite{quim-extremal}
or extended PBH lifetimes via the
``memory burden effect'' \cite{memory-burden,andrea-memory,memory-2}.

In the higher-dimensional case,
and in the absence of further beyond-the-Standard-Model (BSM) physics,
Hawking emission rates for
photons and neutrinos
are expected to be comparable. Ref.~\cite{higher-d-km3}
considered a LED framework in which, motivated by Swampland principles,
the extra-dimensional states form an explicit tower
of sterile neutrinos \cite{dark-dimension}.
This yields an enhanced neutrino flux with a photon--neutrino
branching ratio of 1:6. As explained in Ref.~\cite{higher-d-km3},
supersymmetric extensions could increase this ratio.

Note that an increased neutrino flux is not
an intrinsic feature of higher-dimensional black holes,
but requires BSM physics.
For standard evaporation in higher-dimensional frameworks,
we expect a comparable photon flux and hence $\gamma$-ray signal.

As discussed in Section \ref{section-CMB},
constraints from energy injection into the medium during
Hawking evaporation can place stringent limits on the abundance of
evaporating black holes. Even if the photon–-neutrino branching ratio
is altered by the introduction of additional neutrino degrees of freedom,
electron production can significantly impact the CMB.
Note that $T^{e^-}_c$ in \eqref{def-fc} is generally
larger and peaks at higher redshift than $T^{\gamma}_c$.
Consequently, CMB constraints on candidate sources
prior to the final burst producing the KM3NeT neutrino
must be evaluated carefully for each PBH mass,
even if the multimessenger bounds are circumvented.
The impact on the CMB could be studied by introducing
BSM branching ratios in the \textsc{DarkAges} package
described in Ref.~\cite{CMB-2}
(alternatvely, Ref.~\cite{Friedlander-I}
for higher-dimensional scenarios).
However, this has not yet been done.

Furthermore, emission into the brane is not suppressed,
contrary to early claims in Ref.~\cite{higher-d-km3}.
As shown by Emparan {\etal},
this misconception arose from the interpretation of brane
fields as bulk fields confined to a limited phase space,
rather than intrinsically four-dimensional~\cite{Emparan-brane}.
Proposed
explanations for the KM3NeT event should not rely on this suppression.

Observationally,
higher-dimensional PBHs differ from conventional
ones primarily in their lighter masses at fixed temperature.
This requires higher number densities for a given PBH fraction
$f_{\rm PBH}$.
Moreover, their longer lifetimes allow PBHs that would have
otherwise evaporated long ago to burst today.
Both features can be of use when the likelihood and energy range
of the event are highly constrained, as for the KM3NeT event.

The KM3NeT detection is very intriguing and has attracted
significant interest from the community.
Future observations will be crucial in determining whether
it has a
cosmogenic origin. If similar events are recorded,
multimessenger physics will play an essential role in
filtering source candidates.
%
%
%
%
\section{Microlensing}\label{section-microlensing}
If a compact object travels across the line of sight between
an observer and a star, when sufficiently aligned,
it will lead to formation of both Einstein and relativistic rings.
These are images sourced by light deflection in the
weak and strong field regimes, respectively.
For a subsolar-mass black hole, the displacement of light
is expected to be unresolved.
However, lensing should still lead to an apparent magnification
of the brightness of the source,
which (in principle) could be monitored.
This could be done
by analysing light curves
over a period of time and looking for a sudden peak in an otherwise steady magnitude profile. 

Images in the strong-deflection limit capture
significant near-horizon effects.
These were studied for braneworld scenarios in
Ref.~\cite{eiroa}.
However,
they are extremely challenging to detect.
Our focus is instead on the weak-deflection regime.
The appropriate description of the lensing process
depends on the relative scale between the curvature radius
of the black hole and the wavelength of the light being deflected.
When these are comparable, the wave nature of light becomes important.
We thus make a distinction between the range of black hole masses for
which the geometric optics approximation is valid,
and those for which we need to take into account the effects of wave optics.

Of the available microlensing surveys, only the Hyper Suprime-Cam (HSC) observations are sensitive to the smallest PBH masses, and these are conventionally used to set the upper bound of the dark matter window.
In this case, the sources are stars from M31,
and the lenses are PBHs in the dark matter halo.
The HSC r-band filter detects light of wavelength
$\lambda\approx 0.6\, \mu\text{m}$,
so wave optics is always important for lenses in the
small black hole limit ($\lambda\sim l>r_0$ for all $M$).
For heavier black holes, one can use the geometric description. 

The metric that should be employed
for computation of microlensing observables in braneworlds
is sometimes a source of confusion.
The scales of microlensing are, by definition,
in the domain of linearized gravity.
Therefore, the underlying geometry in braneworlds is the
Garriga--Tanaka solution, regardless the near-horizon black hole solution.
%
%
\subsection{Large black holes}
Following the full analytical description of weak-deflection
lensing in the Garriga--Tanaka framework developed in
Ref.~\cite{keetonpetters},
the total magnification by a ``large'' black hole in RS-II is
\begin{equation}\label{tot-mag}
  \mu_{\rm tot}^{\rm BW}=\left(1-f(u)\,\varepsilon_l^2\right)\mu_{\rm tot}^{4\rm D}+\mathcal{O}\left(\varepsilon_l^3\right)
  .
\end{equation}
Here, $\mu_{\rm tot}^{4\rm D}$
is the expected result for $(3+1)$-dimensions,
and $\varepsilon_l$
is a parameter depending on the AdS curvature,
\begin{equation}
    \label{eq:varepsilon-l}
    \varepsilon_{l}=\frac{\tan^{-1}(l/d_{\rm L})}{\theta_{\rm E}}\,,
\end{equation}
where $d_L$ is the distance to the lens. The multiplicative factor
encodes corrections due to the higher-dimensional
geometry.
These are determined by $f(u)$, which satisfies
\begin{equation}
    f(u)=\frac{4}{(2+u^2)(4+u^2)}
    ,
    \quad u=\beta/\theta_{\rm E}\,,
\end{equation}
where $\beta$ is the angular separation between the lens and the
source and $\theta_{\rm E}$ the $(3+1)$-dimensional angular Einstein radius.
Because this factor is less than unity,
braneworld effects
\emph{decrease} the total magnification.

Equation~\eqref{tot-mag}
is obtained by making a Taylor expansion of the image positions
in powers of $\varepsilon_l$.
The zeroth-order term yields the usual expression for
the weak-deflection limit of the conventional framework.
Braneworld corrections enter at order $\mathcal{O}(\varepsilon_l^2)$.
For the geometrical configuration of the HSC survey,
we have
\begin{equation}
\varepsilon_{l}\approx 10^{8}\left(\frac{10^{-8}M_\odot}{M}\right)^{1/2}\tan^{-1}\left(\frac{l}{10^{21}\rm m}\right)
\end{equation}

Note that the prefactor of the $\varepsilon_l^2$ term
in \eqref{tot-mag}
lies in the range (0, 0.5). We see that the correction depends heavily 
on the ratio of the AdS radius and the distance of the lens.
It follows that,
in any survey covering astrophysical scales,
the correction to the standard result is always negligible,
regardless the black hole mass (within the geometric optics approximation).
We conclude that black holes with masses that exceed the small black hole
limit are expected to be constrained in the same way as in a
$(3+1)$-dimensional scenario.

The first HSC analysis for the conventional scenario was carried out by
Niikura et al.~in 2017 \cite{niikura}. This was followed by
an analysis by Smyth {\etal}
in 2020 \cite{Smyth},
where a more realistic mass distribution of the source stars
in M31 was considered. We use the results in Ref.~\cite{Smyth} for our plots.
%
%
\subsection{Small black holes}
For the ``small'' black hole limit, we must study wave optics effects
in the braneworld framework. These were presented in
Ref.~\cite{keetonpetters},
where the phenomenon of \textit{attolensing} was introduced for the
first time. Like femtolensing in the case of
standard gravity~\cite{femtolensing},
one expects to see interference fringes in the frequency spectrum
once wave optics becomes non-negligible.

In the standard scenario,
femtolensing was discarded as a constraint for primordial black holes
due to arguments presented in \cite{anti-femto1,anti-femto2}.
This applies equally to attolensing.
Interference patterns are observable only if the star source size
projected onto the lens plane is not much larger than the Einstein
radius of the black hole. In the HSC setup, this is not the case
for any lens in the small black hole limit.
Not only are the interference patterns undetectable,
but so is any sort of signal magnification.
Small black holes are thus undetectable in this astrophysical setup.

If a different detection setup were considered where the
finite source size effect is subdominant,
these interference fringes would be hints of braneworld cosmology,
since the patterns predicted in Refs.~\cite{keetonpetters,femtolensing}
are notably different.
The setup proposed in Ref.~\cite{anti-femto2},
where a denser cadence ($10$s)
and g-band monitoring for a sample of white dwarfs over a
year timescale is considered,
would most likely discriminate between models
if interference fringes were detected.
%
%
%
%
%
%
%
%
%
%
%
%
%
\section{Overview and conclusions}\label{section-conclusions}
%
%
%
%
In this work we have investigated the phenomenology of
primordial black holes in Randall--Sundrum Type-II braneworld scenarios,
with particular emphasis on their role as dark matter candidates.
Building on earlier analytic estimates that established the
main qualitative features of primordial black holes in warped geometries,
we have revisited the problem with a more detailed treatment of evaporation
and cosmological evolution.
We have also incorporated
improved astrophysical and cosmological constraints that are
now available. 

At first sight, the restriction of the AdS curvature radius to sub-micrometer values from table-top experiments might seem like a very strong constraint when expressed in everyday or astrophysical units, but in the context of the primordial black hole dark matter window such scales are still comparatively large. This allows a large parameter space with cosmological and astrophysical signatures that can differ substantially from those expected in the conventional flat (3+1)-dimensional scenario. These differences depend greatly on the mass of the black hole. 

A particularly interesting feature of these models is that for
sufficiently large AdS curvature radii, very light black holes---%
with masses as low as $M\sim 10^9\,\text{g}$ for $l\sim 10^{-6}\,\text{m}$---%
could survive until today.
This stands in contrast to the conventional four-dimensional case,
where such black holes would have evaporated long ago unless additional
effects (e.g.~memory burden) are considered. Moreover, their
emission spectra are qualitatively distinct, and could serve as a
discriminant
between the conventional scenario and higher-dimensional
frameworks if a small black hole were to be detected. 

\begin{figure}[htbp]
    \centering
\includegraphics[width=0.6\linewidth]{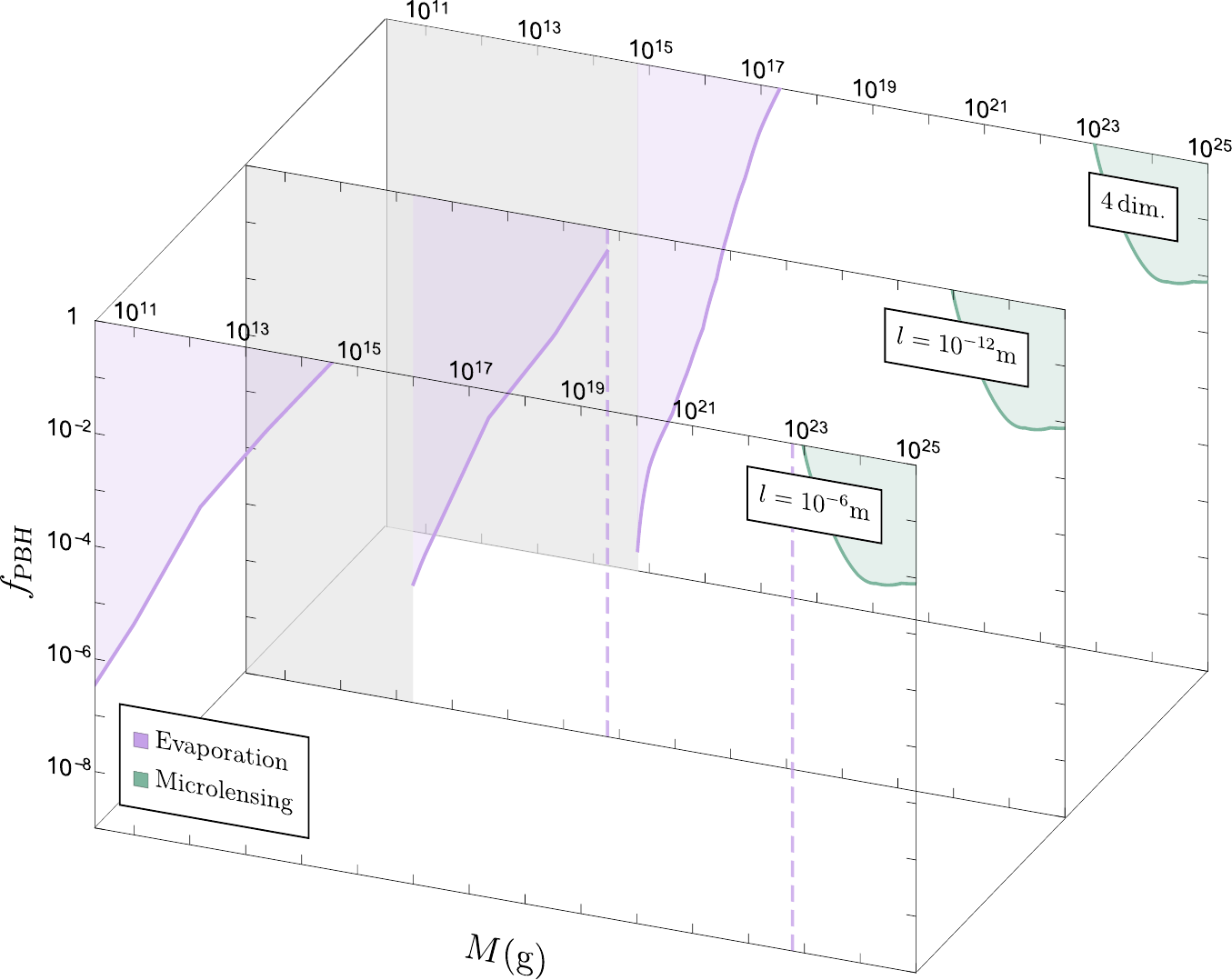}
    \caption{Dark matter window for $l=10^{-6}\,,\,10^{-12}$m and the conventional scenario \cite{Carr-2020} for a monochromatic PBH distribution. Grey--shaded regions indicate black holes that have already evaporated or are currently rapidly evaporating;
    purple envelopes are ($f_{\rm PBH},M$) regions
    excluded by observations related to Hawking evaporation;
    and the green region by the HSC microlensing survey \cite{Smyth}.
    The dashed purple line indicates the limit where $r_0\sim l$ for each particular $l$ choice, i.e.~only black holes lighter than this threshold can be described by the Schwarzschild--Tangherlini solution.}
    \label{final-plot}
\end{figure}

Due to their lower effective temperatures, higher-dimensional
black holes can either evade some observational bounds
altogether, or significantly reduce their severity,
compared to four-dimensional models.
This is particularly true for bounds associated with particle
emission.
Figure~\ref{final-plot} shows that this broadens the allowed
dark matter window by up to about three orders of magnitude. For context, the width is about six orders of magnitude in the conventional scenario.
Similar observations that a non-conventional geometry
can
significantly change the PBH dark matter parameter space
have been made before; see Refs.~\cite{Johnson, Friedlander-I}.
In the particular case of RS-II, the enlargement of the DM window
is more pronounced for larger AdS radii. However, bounds obtained
from X-ray detections remain especially significant in this regime
(see Figure \ref{fig-3d-rad}). Consequently,
the true parameter space is narrower than one might na\"{\i}vely
infer by simply relaxing the most stringent bounds
from the conventional four-dimensional scenario. 

For AdS radii in the range
$10^{-6}\,\text{m} \lesssim l \lesssim 10^{-11}\,\text{m}$,
we find that dark matter could (in principle) be composed
entirely of higher-dimensional black holes.
For smaller curvature scales,
evaporation constraints rule out this possibility.
Conversely, the case of $l\sim 10^{-6}\,\text{m}$ represents an
extreme situation in which the dark matter window is
almost exclusively higher-dimensional. 

What are the prospects for future PBH detection, and especially
for discriminating between the conventional $(3+1)$-dimensional
model and a braneworld scenario?
This picture here is complex.
Interference patterns produced in microlensing
would serve as a clear discriminant between models.
Although not discussed in this paper,
strong-field lensing is also expected to differ
significantly~\cite{eiroa}.
Unfortunately, as is often the case,
our current observational capabilities impose significant limitations.

We have emphasized the importance of accurately modelling the photon flux.
This is crucial if one aims to obtain a realistic estimate
for astrophysical observables---both in local environments,
and for extragalactic surveys. However,
it should be understood that much of our analysis is still
(semi)-analytic.
Therefore, our results are still a simplified
approximation in which many intricacies
are absent.
For example, we do not fully capture
the details of interactions in the interstellar medium,
or the effect of magnetic fields.
We have also assumed a monochromatic PBH mass spectrum.
Extending our results to a broader mass distribution
is straightforward in principle.
We have chosen the idealized monochromatic scenario for
the purpose of maintaining a manageable parameter
space, and to aid visualization.
Similar remarks apply to black hole spin, which we have taken to be
zero.
In the standard formation scenario we do not expect
significant angular momentum.
However, in more exotic formation mechanisms,
such as particle collisions in the early universe,
the spin distribution may be more complicated.
We defer all these issues to future work.
%
%
%
\section*{Acknowledgements}
IAT thanks A.~Reeves, J.~Salvadó and T.~Bertólez-Martínez for very insightful discussions on phenomenology and R.~Emparan for his generous guidance through the intricate concepts and literature surrounding braneworlds and higher-dimensional black holes. IAT also thanks the Institute of Cosmos Sciences of the University of Barcelona (ICCUB) for their hospitality during part of the realization of this project. The research of IAT is supported by an STFC studentship. CB and DS are both supported by the STFC
grants ST/X001040/1 and ST/X000796/1.
%
%
%
\newpage
%
%
%
%
%
%
%
%
%
%


\newpage

\begin{thebibliography}{99}
%
\bibitem{Rizzo}
T.~G.~Rizzo, ``\textit{Pedagogical Introduction to Extra Dimensions,}" SLAC Summer Institute, SLACPUB-10753 (2004)
\href{https://arxiv.org/pdf/hep-ph/0409309v2}{[hep-ph/0409309v2]}.
%
%
\bibitem{review-pbh-1}
A.~M.~Green and B.~J.~Kavanagh, ``\textit{Primordial Black Holes as a dark matter candidate,}" J.~Phys.~G 48 (2021) no.4 \href{https://arxiv.org/abs/2007.10722}{[astro-ph/2007.10722]}.
%
%
\bibitem{review-pbh-2}
A.~Escriv\`{a}, F.~K\"{u}hnel and Y.~Tada, ``\textit{Primordial Black Holes,}" in M. Arca Sedda, E.~Bortolas and M.~Spera (Eds.), ``\textit{Black Holes in the Era of Gravitational-Wave Astronomy}" (pp. 261–377) Elsevier (2024) \href{https://arxiv.org/abs/2211.05767}{[astro-ph/2211.05767]}.
%
%
\bibitem{review-pbh-3}
C.~Byrnes, G.~Franciolini, T.~Harada, P.~Pani and M.~Sasaki, ``\textit{Primordial Black Holes,}" Springer (2025), ISBN 978-981--978886-6, 978-981--978889-7, 978-981--978887-3 doi:10.1007/978-981-97-8887-3.
%
%
\bibitem{PBH-Hawking}
S.~Hawking, ``\textit{Gravitationally Collapsed Objects of Very Low Mass,}" Mon.~Not.~Roy.~Astron.~Soc.~152, 75 (1971).
%
%
\bibitem{LED}
N.~Arkani-Hamed, S.~Dimopoulos and G.~Dvali, ``\textit{The Hierarchy problem and new dimensions at a millimeter,}" 
Phys.~Lett.~B 429, 263–272 (1998) [\href{https://arxiv.org/abs/hep-ph/9803315}{hep-ph/9803315}].
%
%
\bibitem{Friedlander-I}
A.~Friedlander, K.~J.~Mack, S.~Schon, N.~Song and A.~C.~Vincent, ``\textit{Primordial black hole dark matter in the context of extra dimensions,}" Phys.~Rev.~D 105, (2022) 103508 \href{https://arxiv.org/pdf/2201.11761}{[hep-ph/2201.11761]}.
%
%
\bibitem{Friedlander-II}
A.~Friedlander, N.~Song and A.~C.~Vincent, ``\textit{Dark matter from higher-dimensional primordial black holes,}” Phys.~Rev.~D 108 no. 4, (2023) 043523 \href{https://arxiv.org/pdf/2306.01520}{[hep-ph/2306.01520]}.
%
%
\bibitem{Johnson}
G.~Johnson, ``\textit{Primordial Black Hole Constraints with Large Extra Dimensions,}" JCAP 09, 046 (2020) \href{https://arxiv.org/abs/2005.07467}{[astro-ph/2005.07467]}.
%
%
\bibitem{RS-I}
L.~Randall and R.~Sundrum, ``\textit{A Large Mass Hierarchy from a Small Extra Dimension,}" Phys.~Rev.~Lett.~83, 3370 (1999) \href{https://arxiv.org/abs/hep-ph/9905221}{[hep-ph/9905221]}.
%
%
\bibitem{RS-II}
L.~Randall and R.~Sundrum, ``\textit{An Alternative to Compactification,}" Phys.~Rev.~Lett.~83, 4690 (1999) \href{https://arxiv.org/abs/hep-th/9906064}{[hep-ph/9906064]}.
%
%
\bibitem{Clancy-I}
D.~Clancy, R.~Guedens and A.~R.~Liddle. ``\textit{Primordial black holes in braneworld
cosmologies: Formation, cosmological evolution and evaporation}," Phys.~Rev.~D 66 (2002) \href{https://arxiv.org/abs/astro-ph/0205149}{[astro-ph/0205149]}.
%
%
\bibitem{Clancy-II}
R.~Guedens, D.~Clancy and A.~R.~Liddle, ``\textit{Primordial black holes in braneworld cosmologies:
Accretion after formation,}" Phys.~Rev.~D 66 (2002) 083509
\href{https://arxiv.org/abs/astro-ph/0208299}{[astro-ph/0208299]}.
%
%
\bibitem{Majumdar}
 A.~Majumdar, ``\textit{Domination of black hole accretion in brane cosmology,}" Phys.~Rev.~Lett.~90 (3), 031 303 (2003) \href{https://arxiv.org/abs/astro-ph/0208048}{[astro-ph/0208048]}.
%
%
\bibitem{Clancy-III}
D.~Clancy, R.~Guedens and A.~R.~Liddle, ``\textit{Primordial black holes in braneworld cosmologies: astrophysical constraints,}" Phys.~Rev.~D 68, 023507 (2003) \href{https://arxiv.org/abs/astro-ph/0301568}{
[astro-ph/0301568]}.
%
%
\bibitem{Marteens}
 R.~Maartens, ``\textit{Brane World Gravity,}" Living Rev.~Rel.~7, 7 (2004) \href{https://arxiv.org/pdf/gr-qc/0312059}{[gr-qc/0312059]}.
%
%
\bibitem{Emparan}
R.~Emparan and H.~S.~Reall, ``\textit{Black Holes in Higher Dimensions,}" Living Rev.~Rel.~11, 6 (2008) \href{https://arxiv.org/pdf/0801.3471}{[hep-th/0801.3471]}.
%
%
\bibitem{eemeli}
E.~Tomberg, ``\textit{Unit conversions and collected numbers in cosmology, }" (2021) \href{https://arxiv.org/abs/2110.12251}{[astro-ph/2110.12251]}. 
%
%
\bibitem{grav-tests}
 D.~J.~Kapner, T.~S.~Cook, E.~G.~Adelberger, J.~H.~Gundlach, B.~R.~Heckel, C.~D.~Hoyle and H.~E.~Swanson ``\textit{Tests of the gravitational inverse-square law below the dark-energy length scale,}"
Phys.~Rev.~Lett.~98 (2007) 021101 \href{https://arxiv.org/abs/hep-ph/0611184}{[hep-ph/0611184]}.
%
%
\bibitem{KK-BBN-1}
N.~Sasankan, M.~Gangopadhyay, G.~J.~Mathews, and M.~Kusakabe, ``\textit{New observational limits on dark radiation in brane-world cosmology,}" Phys.~Rev.~D 95 (8), 083516 (2017) \href{https://arxiv.org/abs/1607.06858}{[astro-ph/1607.06858]}.
%
%
\bibitem{KK-BBN-2}
N.~Sasankan, G.~J.~Mathews, M.~Kusakabe, and T.~Kajino, ``\textit{Limits on brane-world and particle dark radiation from Big Bang nucleosynthesis and the CMB,}" International Journal of Modern Physics E 26(08), 1741007 (2017) [\href{https://arxiv.org/abs/1706.03630}{astro-ph/1706.03630}].
%
%
\bibitem{ST-metric}
F.~R.~Tangherlini, ``\textit{Schwarzschild field in $n$ dimensions
and the dimensionality of space problem,}" Nuovo Cim. 27
(1963), 636-651.
%
%
\bibitem{fraser}
C.~M.~Fraser and D.~M.~Eardley, ``\textit{The binding energy of small black holes on the brane,}" Class.~Quantum Grav.~32, 055010 (2015) \href{https://arxiv.org/abs/1409.0884}{[gr-qc/1409.0884]}.

\bibitem{frolov-1}
V.~P.~Frolov and D.~Stojković, ``\textit{Black hole as a point radiator and recoil effect on the brane world,}" Phys.~Rev.~Lett.~89, 151302 (2002) \href{https://arxiv.org/abs/hep-th/0208102}{[hep-th/0208102]}.
%
%
\bibitem{frolov-2}
V.~P.~Frolov and D.~Stojković, ``\textit{Black hole radiation in the brane world and recoil effect,}" Phys.~Rev.~D 66, 084002 (2002) \href{https://arxiv.org/abs/hep-th/0206046}{[hep-th/0206046]}.
%
%
\bibitem{flachi}
A.~Flachi and T.~Tanaka, ``\textit{Escape of black holes from the brane,}" Phys.~Rev.~Lett.~95, 161302 (2005) \href{https://arxiv.org/abs/hep-th/0506145}{[hep-th/0506145]}.
%
%
\bibitem{vilenkin}
I.~Olasagasti and A.~Vilenkin, ``\textit{Gravity of higher-dimensional global defects,}" Phys.~Rev.~D 62, 044014 (2000) \href{https://arxiv.org/abs/hep-th/0003300}{[hep-th/0003300]}.
%
%
\bibitem{kaloper}
N.~Kaloper and D.~Kiley, ``\textit{Exact Black Holes and Gravitational Shockwaves on Codimension-2 Branes,}" Phys.~Rev.~D 74, 123509 (2006) \href{https://arxiv.org/pdf/hep-th/0601110}{[hep-th/0601110]}.
%
%
\bibitem{emparan-2}
R.~Emparan, G.~T.~Horowitz and R.~C.~Myers, ``\textit{Exact Description of Black Holes on Branes,}" JHEP
0001, 007 (2000)  \href{https://arxiv.org/abs/hep-th/9911043}{[hep-th/9911043]}.
%
%
\bibitem{Garriga--Tanaka}
J.~Garriga and T.~Tanaka, ``\textit{Gravity in the brane world,}" Phys.~Rev.~Lett.~84 (2000) 2778
\href{https://arxiv.org/abs/hep-th/9911055}{[hep-th/9911055]}.
%
%
\bibitem{Emparan-brane}
R.~Emparan, G.~T.~Horowitz and R.~C.~Myers, ``\textit{Black holes radiate mainly on the brane,}" Phys.~Rev.~Lett.~85 (2000) 499 \href{https://arxiv.org/abs/hep-th/0003118}{[hep-th/0003118]}.
%
%
\bibitem{monochromatic-or-not}
B.~Carr, M.~Raidal, T.~Tenkanen, V.~Vaskonen and H.~Veerm\"{a}e, ``\textit{Primordial black hole constraints for extended mass functions,}" Phys.~Rev.~D 96 (2017) 23514 \href{https://arxiv.org/abs/1705.05567}{[astro-ph/1705.05567]}.
%
%
\bibitem{Gorton:2024cdm}
M.~Gorton and A.~M.~Green, ``\textit{How open is the asteroid-mass primordial black hole window?,}" SciPost Phys.~17 (2024) 2, 032 \href{https://arxiv.org/abs/2403.03839}{[astro-ph/2403.03839]}.
%
%
\bibitem{emparan-3}
R.~Emparan, J.~García-Bellido and N.~Kaloper, ``\textit{Black hole astrophysics in AdS braneworlds,}” JHEP 01 (2003) 079, \href{https://arxiv.org/abs/hep-th/0212132}{[hep-th/0212132]}.
%
%
\bibitem{wiseman}
A.~L.~Fitzpatrick, L.~Randall and T.~Wiseman, ``\textit{On the existence and dynamics of braneworld black holes,}" JHEP 11 (2006) 033 \href{https://arxiv.org/abs/hep-th/0608208}{[hep-th/0608208]}.
%
%
\bibitem{figueras}
P.~Figueras and T.~Wiseman, ``\textit{Gravity and large black holes in Randall–Sundrum II braneworlds,}" Phys.~Rev.~Lett.~107 (2011) 081101 \href{https://arxiv.org/abs/1105.2558}{[hep-th/1105.2558]}.
%
%
\bibitem{emparan-4}
R.~Emparan, R.~Luna, R.~Suzuki, M.~Tomašević and B.~Way, ``\textit{Holographic duals of evaporating black holes,}" JHEP 05 (2023) 182 \href{https://arxiv.org/abs/2301.02587}{[hep-th/2301.02587]}.
%
%
\bibitem{Page-I}
D.~N.~Page, ``\textit{Particle Emission Rates from a Black Hole: Massless Particles from an Uncharged,
 Nonrotating Hole,}" Phys.~Rev.~D 13 (1976) 198.
%
%
\bibitem{kanti-1}
 C.~M.~Harris and P.~Kanti, ``\textit{Hawking radiation from a (4 + n)-dimensional black hole: Exact results for the Schwarzschild phase,}" JHEP 10 (2003) 014 \href{https://arxiv.org/abs/hep-ph/0309054}{[hep-ph/0309054]}.
 %
 %
\bibitem{kanti-2}
 C.~M.~Harris and P.~Kanti, ``\textit{Hawking radiation from a (4 + n)-dimensional rotating black hole,}" Phys.~Lett.~B 633 (2006) 106–110 \href{https://arxiv.org/abs/hep-th/0503010}{[hep-th/0503010]}.
%
%
\bibitem{kanti-3}
M.~Casals, P.~Kanti and E.~Winstanley, ``\textit{Brane decay of a (4 + n)-dimensional rotating black hole. II: Spin-1 particles,}" JHEP 02 (2006) 051 \href{https://arxiv.org/abs/hep-th/0511163}{[hep-th/0511163]}.
%
%
\bibitem{Ida}
 D.~Ida, K.~Y.~Oda and S.~C.~Park, ``\textit{Rotating Black Holes at Future Colliders. III. Determination of Black Hole Evolution,}" Phys.~Rev.~D 73, 124022 (2006) 
\href{https://arxiv.org/pdf/hep-th/0602188}{[hep-th/0602188]}.
%
%
\bibitem{BlackMax}
 D.~C.~Dai, G.~Starkman, D.~Stojkovic, C.~Issever, E.~Rizvi and J.~Tseng, ``\textit{BlackMax: A black-hole event generator with rotation, recoil, split branes and brane tension,}" Phys.~Rev.~D 77
(2008) 076007 \href{https://arxiv.org/abs/0711.3012}{[hep-ph/0711.3012]}.
%
%
\bibitem{tanaka}
N.~Tanahashi and T.~Tanaka, ``\textit{Time-symmetric initial data of large brane-localized black hole in RS-II model,}" Phys.~Rev.~D 77, 084027 (2008) \href{https://arxiv.org/pdf/0712.3799}{[gr-qc/0712.3799]}.
%
%
\bibitem{santos}
W.~D.~Biggs and J.~E.~Santos, ``\textit{Rotating Black Holes in Randall–Sundrum II Braneworlds,}" Phys.~Rev.~Lett.~128, 021601 (2022) \href{https://arxiv.org/abs/2108.00016}{[hep-th/2108.00016]}.
%
%
\bibitem{Carr-2020}
B.~Carr, K.~Kohri, Y.~Sendouda and J.~Yokoyama, ``\textit{Constraints on primordial black holes,}" Rept.~Prog.~Phys.~84, 116902
(2021) \href{https://arxiv.org/pdf/2002.12778}{[astro-ph/2002.12778]}.
%
%
\bibitem{medium-brem}
M.~S.~Longair, ``\textit{High Energy Astrophysics,}" Vol.~I, 2nd Ed.~(Cambridge, 1992).
\\A.~W.~Strong and I.~V.~Moskalenko, ``\textit{Propagation of cosmic-ray nucleons in the galaxy,}" Astrophys.~J.~509 (1998) 212 \href{https://arxiv.org/abs/astro-ph/9807150}{[astro-ph/9807150]}.
\\O.~Lahav and A.~Liddle, ``\textit{The Cosmological Parameters,}" Phys. Lett. B592, 1 (2004) \href{https://arxiv.org/abs/astro-ph/0406681}{[astro-ph/0406681]}.
%
%
\bibitem{Carr-2}
B.~Carr, F. K\"{u}hnel and M. Sandstad, ``\textit{Constraints on primordial black holes from Galactic gamma-ray background,}" Phys.~Rev.~D 94,
no.8, 083504 (2016)
\href{https://arxiv.org/pdf/1604.05349}{[astro-ph/1604.05349]}. 
%
%
\bibitem{Altarelli-Parisi}
Y.~L.~Dokshitzer, ``\textit{Calculation of the Structure Functions for Deep Inelastic Scattering and $e^+e^-$ Annihilation by Perturbation Theory in Quantum Chromodynamics,"} Sov.~Phys.~JETP 46, 641 (1977).
\\V.~N.~Gribov and L.~N.~Lipatov, ``\textit{Deep Inelastic $e^-\,p$ Scattering in Perturbation Theory,}" Sov.~J.~Nucl.~Phys.~15, 438 (1972).
\\G.~Altarelli and G.~Parisi, ``\textit{Asymptotic Freedom in
Parton Language,}" Nucl.~Phys.~B, 126:298–318 (1977).
%
%
\bibitem{splitting-functions}
J.~Chen, T.~Han and B.~Tweedie, ``\textit{Electroweak Splitting Functions and High Energy Showering,}" JHEP, 11:093 (2017) \href{https://arxiv.org/abs/1611.00788}{[hep-th/1611.00788]}.
%
%
\bibitem{FSR-percent}
 R.~Plestid and B.~Zhou, ``\textit{Final state radiation from high and ultrahigh energy neutrino interactions,}" Phys.~Rev.~D 111, 043007 (2025) \href{https://arxiv.org/pdf/2403.07984}{[hep-ph/2403.07984]}.
%
%
\bibitem{Bethe-Bloch}
 H.~A.~Bethe, ``\textit{Zur Theorie des Durchgangs schneller Korpuskularstrahlen durch Materie,}" Ann.~Phys.~397, 325
(1930).
\\H.~A.~Bethe, ``\textit{Bremsformel für Elektronen relativistischer
Geschwindigkeit,}" Z.~Phys.~76, 293 (1932).
\\H.~Bethe and J.~Ashkin, ``\textit{Experimental Nuclear Physics,}" J.~Wiley, New York, 253 (1953). 
%
%
\bibitem{Beacom}
J.~F.~Beacom and H.~Y\"{u}ksel, ``\textit{Stringent Constraint
on Galactic Positron Production,}" Phys.~Rev.~Lett.~97,
071102 (2006) \href{https://arxiv.org/pdf/astro-ph/0512411}{[astro-ph/0512411]}.
%
%
\bibitem{Jean-2009}
P.~Jean, W.~Gillard, A.~Marcowith and K.~Ferriere, ``\textit{Positron transport in the interstellar medium,}" A\&A 508,
1099 (2009) \href{https://arxiv.org/pdf/0909.4022}{[astro-ph/0909.4022]}.
%
%
\bibitem{0007032}
R.~J.~Gould and E.~H.~Liang, ``\textit{Positron Annihilation in the Galaxy,}" (2000) \href{https://arxiv.org/abs/astro-ph/0007032}{[astro-ph/0007032]}.
%
%
\bibitem{ann-cross}
J.~M.~Jauch and F.~Rohrlich, ``\textit{The Theory of Photons and Electrons: The Relativistic Quantum Field Theory of Charged Particles with Spin One-half}", (2nd ed.) Springer (1976).
%
%
\bibitem{fIA}
A.~A.~Zdziarski and R.~Svensson, ``\textit{Inverse Compton emission from electron–positron pairs}", Astrophys.~J.~344, 551 (2000).  
%
%
\bibitem{Boehm}
N.~Prantzos, C.~Boehm, A.~M.~Bykov \textit{et al.}, ``\textit{The 511 keV emission from positron annihilation in the Galaxy,}" Reviews of Modern Physics, 83, 1001 (2011) \href{https://arxiv.org/abs/1009.4620}{[astro-ph/1009.4620]}.
%
%
\bibitem{positronium}
T.~Siegert, R.~Diehl, G.~Khachatryan, M.~G.~H.~Krause, F.~Guglielmetti, J.~Greiner, A.~W.~Strong and X.~Zhang, \textit{``Gamma-ray spectroscopy of Positron Annihilation in the Milky Way,}" Astron.~Astrophys.~586 (2016) \href{https://arxiv.org/abs/1512.00325}{[astro-ph/1512.00325]}.
%
%
\bibitem{3-insteadof-2}
A.~Ore and J.~L.~Powell, ``\textit{Three-Photon Annihilation of an Electron-Positron Pair,}" Phys.~Rev.~75, 1696 (1949).
%
%
\bibitem{Femenia}
 P.~D.~Ruiz-Femenia, ``\textit{Orthopositronium decay spectrum using NRQED,}" Nucl.~Phys.~Proc.~Suppl.~152 (2006) 200 \href{https://arxiv.org/pdf/hep-ph/0311002}{[hep-ph/0311002]}.
 %
 %
\bibitem{NFW}
J.~F.~Navarro, C.~S.~Frenk and S.~D.~M.~White, ``\textit{The Structure of Cold Dark Matter Halos,}" Astrophys.~J.~462, 563 (1996) \href{https://arxiv.org/abs/astro-ph/9508025}{[arXiv:astro-ph/9508025]}.
\\J.~F.~Navarro, C.~S.~Frenk and S.~D.~M.~White, ``\textit{A Universal Density Profile from Hierarchical Clustering,}" Astrophys.~J.~490, 493 (1997) \href{https://arxiv.org/abs/astro-ph/9611107}{[astro-ph/9611107]}.
%
%
 \bibitem{NFW-parameters}
M.~Cautun \textit{et al.}, ``\textit{The Milky Way total mass profile as inferred from Gaia DR2,}” (2019) \href{https://arxiv.org/pdf/1911.04557}{[astro-ph/1911.04557]}.
%
%
\bibitem{Tinker}
J.~Tinker, A.~V.~Kravtsov, A.~Klypin, K.~Abazajian,
M.~Warren, G.~Yepes, S.~Gottl\"{o}ber and D.~E.~Holz, ``\textit{Toward a halo mass function for precision cosmology:
The limits of universality,}" Astrophys.~J.~688, 709–728 (2008) \href{https://arxiv.org/abs/0803.2706}{[astro-ph/0803.2706]}.
%
%
\bibitem{comparison-HMF}
W.~Chui, ``\textit{How much do we know the halo mass function? Predictions beyond resolution,}" (2024) \href{https://arxiv.org/html/2406.03829v1}{[astro-ph/2406.03829v1]}.
%
%
\bibitem{PS-HMF}
W.~H.~Press and P.~Schechter, ``\textit{Formation of galaxies and clusters of galaxies by selfsimilar gravitational condensation,}" Astrophys.~J.~187 (1974) 425-438.
%
%
\bibitem{DM-halos-baryons}
H.~Zheng, S.~Bose, C.~S.~Frenk, L.~Gao, A.~Jenkins \textit{et al.}, ``\textit{The influence of baryons on low-mass haloes,}"  Mon.~Not.~Roy.~Astron.~Soc.~532 (2024) 3, 3151-3165 \href{https://arxiv.org/pdf/2403.17044}{[astro-ph/2403.17044]}.
%
%
\bibitem{baryon-fraction}
R.~J.~Wright, R.~S.~Somerville, C.~del~P.~Lagos, M.~Schaller, R.~Davé, D.~Anglés-Alcázar and S.~Genel, ``\textit{The baryon cycle in modern cosmological hydrodynamical simulations,}"  Accepted for publication in MNRAS (2024) \href{https://arxiv.org/abs/2402.08408}{[astro-ph/2402.08408]}.
%
%
\bibitem{DESI}
DESI Collaboration \textit{et al.} (2024)
\href{https://arxiv.org/pdf/2404.03002}{[astro-ph/2404.03002]}.
%
%
\bibitem{Hooper-2}
C.~Blanco and D.~Hooper, ``\textit{Constraints on Decaying Dark
Matter from the Isotropic Gamma-Ray Background,}" JCAP
03 (2019) 019 \href{https://arxiv.org/pdf/1811.05988}{[astro-ph/1811.05988]}.
%
%
\bibitem{data-GRE}
L.~Bouchet, A.~W.~Strong, T.~A.~Porter, I.~V.~Moskalenko, E.~Jourdain and J.~P.~Roques, ``\textit{Diffuse emission measurement with INTEGRAL/SPI as indirect probe of
cosmic-ray electrons and positrons,}"
Astrophys.~J.~739, 29 (2011)  
\href{https://arxiv.org/pdf/1107.0200}{[astro-ph/1107.0200]}.
%
%
\bibitem{Laha-GRE}
R.~Laha, J.~B.~Mu\~noz and T.~R.~Slatyer, ``\textit{INTEGRAL constraints on primordial black holes and particle dark matter,}"
Phys.~Rev.~D 101 (2020) 123514 \href{https://arxiv.org/pdf/2004.00627}{[astro-ph/2004.00627]}.
%
%
\bibitem{sterile-neutrino}
D.~Sicilian, N.~Cappelluti, E.~Bulbul, F.~Civano, M.~Moscetti and C.~S.~Reynolds, ``\textit{Probing the milky way’s dark
matter halo for the 3.5 keV line,}" Astrophys.~J.~905 (2020) 146
\href{https://arxiv.org/pdf/2008.02283}{[astro-ph/2008.02283]}.
%
%
\bibitem{XMM-Newton-data}
J.~W.~Foster, M.~Kongsore, C.~Dessert, Y.~Park, N.~L.~Rodd, K.~Cranmer \textit{et al.}, ``\textit{Deep search
for decaying dark matter with XMM-newton blank-sky observations,}", Phys.~Rev.~Lett.~127 (2021) 051101 \href{https://arxiv.org/abs/2102.02207}{[astro-ph/2102.02207]}.
%
%
\bibitem{extract-eRos-XMM}
S.~Balaji, D.~Cleaver, P.~De la Torre Luque and M.~Michailidis, ``\textit{Dark Matter in X-rays: Revised XMM-Newton Limits and New Constraints from eROSITA
,}" (2025) \href{https://arxiv.org/abs/2506.02310}{[hep-ph/2506.02310]}.
%
%
\bibitem{cappelluti}
N.~Cappelluti, Y.~Li, A.~Ricarte \textit{et al.}, ``\textit{The Chandra COSMOS legacy survey: Energy Spectrum of the Cosmic X-ray Background and constraints on undetected populations,}" Astrophys.~J.~837, 19 (2017)
\href{https://arxiv.org/pdf/1702.01660}{[astro-ph/1702.01660]}.
%
%
\bibitem{223}
D.~E.~Gruber, J.~L.~Matteson, L.~E.~Peterson and G.~V.~Jung, ``\textit{The Spectrum of Diffuse Cosmic Hard X-Rays Measured with HEAO-1,}" Astrophys.~J.~520, 124 (1999) \href{https://arxiv.org/abs/astro-ph/9903492}{[astro-ph/9903492]}.
%
%
\bibitem{224}
G.~Weidenspointner \textit{et al.}, ``\textit{The Comptel instrumental line background,}"  AIP Conf.~Proc.~510 (2000) 1, 581-585, Astron.~Astrophys.~368 (2001) 347 \href{https://arxiv.org/pdf/astro-ph/0012332}{[astro-ph/0012332]}.
%
%
\bibitem{225}
A.~W.~Strong, I.~V.~Moskalenko and O.~Reimer, ``\textit{A new determination of the extragalactic diffuse gamma-ray background from EGRET data,}"
Astrophys.~J.~613, 956 (2004) \href{https://arxiv.org/abs/astro-ph/0405441}{[astro-ph/0405441]}.
%
%
\bibitem{226}
A.~Abdo et al. (Fermi LAT), ``\textit{The Spectrum of the Isotropic Diffuse Gamma-Ray Emission Derived From First-Year Fermi Large Area Telescope Data,}" Phys.~Rev.~Lett.~104, 101101 (2010) \href{https://arxiv.org/abs/1002.3603}{[astro-ph/1002.3603]}.
%
%
\bibitem{balloon-1972}
W.~N.~Johnson, F.~R.~Harnden and R.~C.~Haymes, ``\textit{Evidence for Hard X-Ray Pulsations from the Vela Pulsar,}" Astrophys.~J.~172, L1 NASA ADS (1972).
%
%
\bibitem{integral-spi}
J.~Knödlseder, V.~Lonjou, G.~Weidenspointner et al., ``\textit{The all-sky distribution of 511 keV electron-positron annihilation emission,}" A\&A, 441, 513 (2005) [\href{https://arxiv.org/abs/astro-ph/0506026}{astro-ph/0506026}].
\\G.~Weidenspointner, C.~R.~Shrader, J.~Knödlseder \textit{et al.}, ``\textit{The sky distribution of positronium annihilation continuum emission measured with SPI/INTEGRAL,}" A\&A, 450, 1013 (2006) [\href{https://arxiv.org/abs/astro-ph/0601673}{astro-ph/0601673}].
\\C.~A.~Kierans \textit{et al.}, ``\textit{Positron Annihilation
in the Galaxy,}" Astrophys.~J.~895, 44 (2020) [\href{https://arxiv.org/abs/1903.05569}{astro-ph/1903.05569}].
%
%
\bibitem{astro-511}
E.~Kalemci, S.~E.~Boggs, P.~A.~Milne and S.~P.~Reynolds, ``\textit{Searching for annihilation radiation from SN 1006 with SPI on INTEGRAL,}" Astrophys.~J.~Lett.~640, L55 (2006) [\href{https://arxiv.org/abs/astro-ph/0602233}{astro-ph/0602233}].
\\M.~Casse, B.~Cordier, J.~Paul and S.~Schanne, ``\textit{Hypernovae/GRB in the Galactic Center as possible sources of Galactic Positrons,}" Astrophys.~J.~Lett.~602, L17 (2004) [\href{https://arxiv.org/abs/astro-ph/0309824}{astro-ph/0309824}].
\\G.~Bertone, A.~Kusenko, S.~Palomares-Ruiz, S.~Pascoli and D.~Semikoz, ``\textit{Gamma ray bursts and the origin of galactic positrons,}" Phys.~Lett.~B~636, 20 (2006) [\href{https://arxiv.org/abs/astro-ph/0405005}{astro-ph/0405005}].
\\N.~Guessoum, P.~Jean and N.~Prantzos, ``\textit{Microquasars as sources of positron annihilation radiation,}" Astron.~Astrophys.~457, 753 (2006)  [\href{https://arxiv.org/abs/astro-ph/0607296}{astro-ph/0607296}].
\\R.~Bartels, F.~Calore, E.~Storm and C.~Weniger, ``\textit{Galactic Binaries Can Explain the Fermi Galactic Center Excess and 511 keV Emission,}" Mon.~Not.~Roy.~Astron.~Soc.~480, 3826 (2018) [\href{https://arxiv.org/abs/1803.04370}{astro-ph/1803.04370}].
\\V.~Takhistov, ``\textit{Novel Signals from Neutron Star Mergers at 511 keV,}" PoS ICRC2019, 803 (2020) [\href{https://arxiv.org/abs/1908.01100}{astro-ph/1908.01100}].
\\G.~M.~Fuller, A.~Kusenko, D.~Radice and V.~Takhistov, ``\textit{Positrons and 511 keV radiation as tracers of recent binary neutron star mergers,}" Phys.~Rev.~Lett.~122, 121101 (2019) [\href{https://arxiv.org/abs/1811.00133}{astro-ph/1811.00133}].
%
%
\bibitem{review-511}
N.~Prantzos \textit{et al.}, ``\textit{The 511 keV emission from positron annihilation in the Galaxy,}" Rev.~Mod.~Phys.~83, 1001 (2011) [\href{https://arxiv.org/abs/1009.4620}{astro-ph/1009.4620}].
%
%
\bibitem{lightDM-511}
Y.~Ascasibar, P.~Jean, C.~Boehm and J.~Knoedlseder, ``\textit{Constraints on dark matter and the shape of the Milky Way dark halo from the 511 keV line,}" Mon.~Not.~Roy.~Astron.~Soc.~368:1695-1705 (2006) [\href{https://arxiv.org/abs/astro-ph/0507142}{astro-ph/0507142}].
\\Y.~Rasera, R.~Teyssier, P.~Sizun, M.~Casse, P.~Fayet, B.~Cordier and J.~Paul, ``\textit{Soft gamma-ray background and light Dark Matter annihilation,}" Phys.~Rev.~D 73 103518 (2006) [\href{https://arxiv.org/abs/astro-ph/0507707}{astro-ph/0507707}].
%
%
\bibitem{okele}
P.~N.~Okele and M.~J.~Rees, ``\textit{ Observational consequences of positron production by evaporating black holes,}" Astron.~Astrophys.~81, 263 (1980).
%
%
\bibitem{ift}
P.~De la Torre Luque, J.~Koechler and S.~Balaji, ``\textit{Refining Galactic primordial black hole
evaporation constraints,}" Phys.~Rev.~D 110 (2024) 123022 \href{https://arxiv.org/pdf/2406.11949}{[astro-ph/2406.11949]}.
%
%
\bibitem{hooper511}
C.~Keith and D.~Hooper, ``\textit{511 keV excess and primordial black holes,}" Phys.~Rev.~D 104, 063033 (2021) \href{https://arxiv.org/pdf/2103.08611}{[astro-ph/2103.08611]}.
%
%
\bibitem{1906.00498}
T.~Siegert, R.~M.~Crocker, R.~Diehl, M.~G.~H.~Krause, F.~H.~Panther, M.~Pleintinger and C.~Weinberger, ``\textit{Constraints on positron annihilation kinematics in the inner Galaxy,}" A\&A 627, A126 (2019) \href{https://arxiv.org/pdf/1906.00498}{[astro-ph/1906.00498]}.
%
%
\bibitem{graham}
W.~DeRocco and P.~W.~Graham, ``\textit{Constraining Primordial Black Hole Abundance with the Galactic 511 keV
Line,}" Phys.~Rev.~Lett.~123, 251102 (2019) \href{https://arxiv.org/abs/1906.07740}{[astro-ph/1906.07740]}.
%
%
\bibitem{voy-pbh-first-paper}
M.~Boudaud and M.~Cirelli, ``\textit{Voyager 1 $e^\pm$ Further Constrain Primordial Black Holes as Dark
Matter,"} Phys.~Rev.~Lett.~122 (2019) 041104 \href{https://arxiv.org/pdf/1807.03075}{[astro-ph/1807.03075]}.
%
%
\bibitem{voy-data}
L.~F.~Burlaga, N.~F.~Ness and
E.~Stone, ``\textit{Magnetic Field Observations as Voyager 1 Entered the Heliosheath Depletion Region,}" Science 341, 147-150 (2013).
\\A.~C.~Cummings, E.~C.~Stone, B.~C.~Heikkila, N.~Lal, W.~R.~Webber, G.~Johannesson, I.~V.~Moskalenko, E.~Orlando and T.~A.~Porter, ``\textit{Galactic Cosmic Rays Throughout the Heliosphere and in the Local Interstellar Medium,}" Astrophys.~J.~831, 18 (2016).
\\E.~Stone, A.~Cummings and B.~Heikkila, ``\textit{Cosmic ray measurements from voyager 2 as it crossed
into interstellar space,}” \href{https://www.nature.com/articles/s41550-019-0928-3}{Nature Astronomy 3 (11, 2019) 1013–1018}.
%
%
\bibitem{nH-helio}
K.~Dialynas, S.~M.~Krimigis, R.~B.~Decker and D.~G.~Mitchell, ``\textit{Plasma pressures in the heliosheath from Cassini ENA and Voyager 2 measurements: Validation by the Voyager 2 heliopause crossing,}" 
\href{https://arxiv.org/pdf/1907.03425}{[physics.space-ph/1907.03425]}.
%
%
\bibitem{ams-data}
AMS Collaboration, M.~Aguilar \textit{et al.}, ``\textit{The Alpha Magnetic Spectrometer (AMS) on the
international space station: Part II — Results from the first seven years,}" Phys.~Rept.~894 (2021) 1–116.
%
%
\bibitem{voy-pbh-second-paper}
J.~Huang and Y.~Zhou, ``\textit{Constraints on evaporating primordial black holes from the AMS-02 positron data,}" accepted for publication in Phys.~Rev.~D (2025) \href{https://arxiv.org/pdf/2403.04987}{[astro-ph/2403.04987]}.
%
%
%
%
\bibitem{CMB-1}
V.~Poulin, J.~Lesgourgues and P.~D.~Serpico, ``\textit{Cosmological constraints on exotic injection of
electromagnetic energy,}" JCAP, vol.~1703, no.~03, p.~043 (2017) \href{https://arxiv.org/pdf/1610.10051}{[astro-ph/1610.10051]}.
%
%
\bibitem{CMB-2}
P.~St\"{o}cker, M.~Kr\"{a}mer, J.~Lesgourgues and V.~Poulin, ``\textit{Exotic energy injection with ExoCLASS:
Application to the Higgs portal model and evaporating black holes,}" JCAP 03 (2018) 018 \href{https://arxiv.org/pdf/1801.01871}{[astro-ph/1801.01871]}.
%
%
\bibitem{cmb-transferfunction1}
T.~R.~Slatyer, ``\textit{Indirect dark matter signatures in the cosmic dark ages. I. Generalizing the bound on s-wave dark matter annihilation from Planck results,}" Phys.~Rev.~D
93, 023527 (2016) \href{https://arxiv.org/abs/1506.03811}{[hep-ph/1506.03811]}.
%
%
\bibitem{cmb-transferfunction2}
T.~R.~Slatyer, ``\textit{Indirect dark matter signatures in the cosmic dark ages II. Ionization, heating and photon production from arbitrary energy injections,}" Phys.~Rev.~D 93, 023521 (2016) \href{https://arxiv.org/abs/1506.03812}{[astro-ph/1506.03812]}.
%
%
\bibitem{CLASS}
D.~Blas, J.~Lesgourgues and T.~Tram, ``\textit{The Cosmic Linear Anisotropy Solving System
(CLASS) II: Approximation schemes,}" JCAP, vol.~1107, p.~034 (2011) \href{https://arxiv.org/abs/1104.2933}{[astro-ph/1104.2933]}.
%
%
\bibitem{higher-d-km3}
L.~A.~Anchordoqui, F.~Halzen and D.~Lüst, ``\textit{Neutrinos from Primordial Black Holes in Theories with Extra Dimensions,}" (2025) \href{https://arxiv.org/pdf/2505.23414}{[hep-ph/2505.23414]}.
%
%
\bibitem{higher-d-km3-2}
L.~A.~Anchordoqui, A.~Bedroya and D.~Lüst,
``\textit{Primordial Black Holes are 5D,}" (2025) \href{https://arxiv.org/abs/2506.14874}{[hep-ph/2506.14874]}.
%
%
\bibitem{km3-detection}
 J.~Coelho, ``\textit{Latest results from KM3NeT,}" in XXXI International
Conference on Neutrino Physics and Astrophysics (2024) on
behalf of the KM3NeT Collaboration.
\\S.~Aiello \textit {et al.} (KM3NeT), ``\textit{Observation of an ultra-high-energy
cosmic neutrino with KM3NeT,}" Nature 638, 376 (2025).
%
%
\bibitem{gamma-1}
A.~Klipfel and D.~I.~Kaiser, ``\textit{Ultra‑High‑Energy Neutrinos from Primordial Black Holes,}" (2025)
\href{https://arxiv.org/pdf/2503.19227}{[hep-ph/2503.19227]}.
%
%
\bibitem{gamma-2}
L.~F.~T.~Airoldi, G.~F.~S.~Alves, Y.~F.~Perez‑Gonzalez, G.~M.~Salla and R.~Zukanovich Funchal, ``\textit{Could a Primordial Black Hole Explosion Explain the KM3NeT Event?,}" (2025)
\href{https://arxiv.org/pdf/2505.24666}{[hep-ph/2505.24666]}.
%
%
\bibitem{km3-icecube}
%
%
\bibitem{km3-icecube-2}
S.~W.~Li, P.~Machado, D.~Naredo-Tuero and T.~Schwemberger, ``\textit{Clash of the Titans: ultra‑high energy KM3NeT event versus IceCube data,}" (2025) \href{https://arxiv.org/pdf/2502.04508}{[astro-ph/2502.04508]}.
%
%
\bibitem{quim-extremal}
M.~J.~Baker, J.~Iguaz Juan, A.~Symons and A.~Thamm, ``\textit{Explaining the PeV Neutrino Fluxes at KM3NeT and IceCube with Quasi‑Extremal Primordial Black Holes,}" (2025) \href{https://arxiv.org/abs/2505.22722}{[hep-ph/2505.22722]}.
%
%
\bibitem{memory-burden}
G.~Dvali, ``\textit{A microscopic model of holography: Survival by the burden of memory,}" \href{https://arxiv.org/abs/1810.02336}{[hep-ph/1810.02336]}. 
\\G.~Dvali, L.~Eisemann, M.~Michel and S.~Zell, ``\textit{Universe’s primordial quantum memories,}" JCAP 03, 010 (2019)
\href{https://arxiv.org/abs/1812.08749}{[hep-th/1812.08749]}.
%
%
\bibitem{andrea-memory}
A.~Boccia and F.~Iocco, ``\textit{A strike of luck: could the KM3‑230213A event be caused by an evaporating primordial black hole?,}" (2025) \href{https://arxiv.org/abs/2502.19245}{[astro-ph/2502.19245]}.
%
%
\bibitem{memory-2}
G.~Dvali, M.~Zantedeschi and S.~Zell, ``\textit{Transitioning to Memory Burden: Detectable Small Primordial Black Holes as Dark Matter,}" (2025) \href{https://arxiv.org/abs/2503.21740}{[hep-ph/2503.21740]}.
%
%
\bibitem{dark-dimension}
M.~Montero, C.~Vafa and I.~Valenzuela, ``\textit{The Dark Dimension and the Swampland,}" JHEP 02 (2023) 022 \href{https://arxiv.org/pdf/2205.12293}{[hep-th/2205.12293]}.
%
%
\bibitem{eiroa}
E.~F.~Eiroa, ``\textit{Braneworld Black Hole Gravitational Lens: Strong Field Limit Analysis,}" Phys.~Rev.~D 71 (2005) 083010 \href{https://arxiv.org/abs/gr-qc/0410128}{[gr-qc/0410128]}.
%
%
\bibitem{keetonpetters}
C.~R.~Keeton and A.~O.~Petters, ``\textit{Formalism for testing theories of gravity using lensing by compact objects. III: Braneworld gravity,}"  Phys.~Rev.~D 73 (2006) \href{https://arxiv.org/abs/gr-qc/0603061}{[gr-qc/0603061]}.
%
%
 \bibitem{niikura}
H.~Niikura, M.~Takada, S.~Yokoyama, T.~Sumi and
S.~Masaki, ``\textit{Microlensing constraints on primordial black holes with the Subaru/HSC Andromeda observation,}" Nature Astronomy (2019) \href{https://arxiv.org/abs/1701.02151}{[astro-ph/1701.02151]}.
%
%
\bibitem{Smyth}
N.~Smyth, S.~Profumo, S.~English, T.~Jeltema, K.~McKinnon and
P.~Guhathakurta, ``\textit{Updated constraints on asteroid-mass
primordial black holes as dark matter,}" Phys. Rev. D 101, 7
063005 (2020) \href{https://arxiv.org/pdf/1910.01285}{[astro-ph/1910.01285]}. 
%
%
\bibitem{femtolensing}
A.~Gould, ``\textit{Femtolensing of Gamma-Ray Bursters
,}" Astrophys.~J.~386, L5 (1992).
%
%
\bibitem{anti-femto1}
A.~Katz, J.~Kopp, S.~Sibiryakov and W.~Xue, ``\textit{Femtolensing by Dark Matter Revisited,}" JCAP
12 (2018) 005 \href{https://arxiv.org/pdf/1807.11495.pdf}{[astro-ph/1807.11495]}.
%
%
\bibitem{anti-femto2}
S.~Sugiyama, T.~Kurita and M.~Takada, ``\textit{On the wave optics effect on primordial black hole
constraints from optical microlensing search,}" Mon.~Not.~Roy.~Astron.~Soc.~493, 3632 (2020) \href{https://arxiv.org/pdf/1905.06066.pdf}{[astro-ph/1905.06066]}.
%
%
\end{thebibliography}
\end{document}